# Interpreting the Ionization Sequence in Star-Forming Galaxy Emission-Line Spectra

Chris T. Richardson,[1*] James T. Allen,[2*] Jack A. Baldwin,[3*] Paul C. Hewett,[4*] Gary J. Ferland,[5*] Anthony Crider,[1*] and Helen Meskhidze[1*]

[1]*Physics Department, Elon University, Elon, NC 27215, USA*
[2]*Sydney Institute for Astronomy, School of Physics, University of Sydney, NSW 2006, Australia*
[3]*Physics & Astronomy Department, Michigan State University, East Lansing, MI 48864-1116, USA*
[4]*Institute of Astronomy, University of Cambridge, Madingley Road, Cambridge CB3 0HA*
[5]*Physics and Astronomy Department, University of Kentucky, Lexington, KY, 40506-0055, USA*

**Abstract**

High ionization star forming (SF) galaxies are easily identified with strong emission line techniques such as the BPT diagram, and form an obvious ionization sequence on such diagrams. We use a locally optimally emitting cloud model to fit emission line ratios that constrain the excitation mechanism, spectral energy distribution, abundances and physical conditions along the star-formation ionization sequence. Our analysis takes advantage of the identification of a sample of pure star-forming galaxies, to define the ionization sequence, via mean field independent component analysis. Previous work has suggested that the major parameter controlling the ionization level in SF galaxies is the metallicity. Here we show that the observed SF-sequence could alternatively be interpreted primarily as a sequence in the distribution of the ionizing flux incident on gas spread throughout a galaxy. Metallicity variations remain necessary to model the SF-sequence, however, our best models indicate that galaxies with the highest and lowest observed ionization levels (outside the range $-0.37 < \log$ [O III]/H$\beta$ $< -0.09$) require the variation of an additional physical parameter other than metallicity, which we determine to be the distribution of ionizing flux in the galaxy.

## 1. Introduction

Starburst galaxies are defined by the vigorous formation of massive stars, leading to an emission-line spectrum that is dominated by gas ionized by O stars recently formed in molecular clouds. Early work by Baldwin, Phillips and Terlevich (1981), later expanded upon by Veilleux and Osterbrock (1987; hereafter VO87), showed that the intensity ratios of strong emission lines (known as "strong-line ratios") found in active galactic nuclei (AGN) and star forming (SF) galaxies can be used to classify the excitation mechanism in emission line galaxies. In particular, the plot of [O III] λ5007/ Hβ vs. [N II] λ6584/ Hα (hereafter the BPT diagram) has been remarkably successful at separating galaxies highly ionized by a central AGN from those ionized by intense starbursts. On the BPT diagram, SF galaxies fall in a well-defined region that forms a

---

* crichardson17@elon.edu (CTR); j.allen@physics.usyd.edu.au (JTA); baldwin@pa.msu.edu (JAB); phewett@ast.cam.ac.uk (PCH); gary@pa.uky.edu (GJF); acrider@elon.edu (AC); emeskhidze@elon.edu (HM)

sequence spanning approximately an order of magnitude in these ionization-sensitive line ratios.

A widely-used model explains this sequence by attributing it mainly to systematic chemical abundance variations (Kewley et al. 2013; 2010; Levesque, Kewley & Larson 2010 (hereafter L10)). The other major variable in this model is the ionization parameter $U = \phi_H/n_H c$, where $\phi_H$ is the ionizing photon flux and $n_H$ is the hydrogen density, which is assumed to have a single value for each galaxy in question but which varies between galaxies. Similar models using the filling factor of the ionized gas as a surrogate for $U$ were described by Moy, Rocca-Volmerange and Fioc (2001). These models can successfully fit the VO87 optical line ratio diagrams over a large range in ionization.

In Richardson et al. (2014; hereafter Paper II), we performed plasma simulations with the spectral synthesis code CLOUDY (Ferland et al. 2013) to reproduce AGN observations (Allen et al. 2013; hereafter Paper I) over a wide range of ionization with a locally optimally emitting cloud (LOC) model (Ferguson et al. 1997). This model assumes that the cumulative observed emission from each individual emission line galaxy is the result of selection effects stemming from various emission lines optimally emitted by a large number of gas clouds spanning a large range in physical conditions. After integrating over clouds with a wide range of gas density and radial distance from the ionizing continuum source, weighting the gas clouds by power laws in these two parameters, we found that almost all of our higher-to-moderate ionization diagnostic line ratios could be represented by a systematic change in the radial integration weighting parameter along the AGN ionization sequence. Physically, this represents a systematic change in the radial distribution of clouds around the source. In several cases, especially for the classical line ratio diagrams, we fit the entire sequence of observations from Seyferts all the way down to low ionization AGN. In addition, we also used weaker lines in diagnostic diagrams to act as consistency checks on our models.

In the present paper, we perform a similar analysis for our subsets of SF galaxies. LOC models have successfully fit observations in the NLR (Ferguson et al. 1997) and BLR (Korista et al. 1997) of AGN and here we extend the use of LOC models to describe emission lines originating from SF galaxies. Fig. 1 illustrates two different morphological types of SF galaxies, which support the use of an LOC model. NGC 4038/4039, commonly known as the Antennae Galaxies, represents a typical starburst galaxy undergoing an anomalously high rate of star formation due to a recent merger. Fig. 1 also shows M33, commonly known as the Triangulum Galaxy, which represents an average spiral galaxy undergoing a typical rate of star formation. The giant H II region NGC 604, which actually lies within the Triangulum Galaxy, is shown to scale against the galaxy indicating that many giant H II regions reside in the spiral arms along with several smaller sized H II regions. Thus, in spite of potentially having very different morphological structures, the composite nebular spectrum originating from star forming galaxies ultimately depends on the cumulative emission coming from H II regions that vary substantially in size.

In contrast to other popular modeling methodologies, an LOC model takes into account the differences within each individual galaxy. Fig. 2 compares the physical representation of a star forming galaxy using the setup outlined in L10 and the setup for a typical LOC model. The shading within each H II region (circle) represents the $\phi_H$ value and the shading in the background represents the $n_H$ value around that H II region. For the L10 setup, the gas within a single galaxy is modeled by assuming one value for $\phi_H$, $n_H$, and $Z$, each characteristic of the values present within galactic H II regions. Thus for a constant spectral energy distribution (SED), the size of each H II region would be similar within a galaxy. For the LOC model setup, gas within a galaxy is modeled by assuming a *distribution* of $\phi_H$ and $n_H$ values along with a single *Z* value. Thus for a constant SED, the size of the H II regions would be variable. This explains why in Fig. 2 the LOC model shows circles of various size, which when compared to the actual observations displayed in Fig. 1, more accurately depicts the physical characteristics of a star forming galaxy.

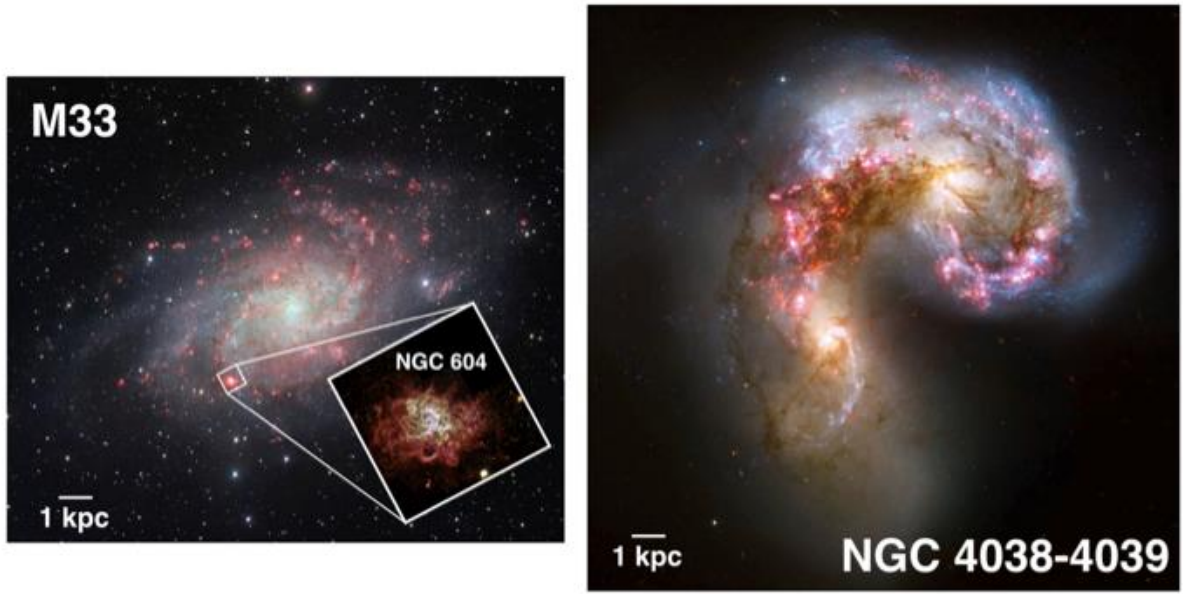

**Figure 1.** The Triangulum Galaxy (M33; Credit: ESO) and the Antennae Galaxies (NGC 4038 and NGC 4039; Credit: NASA, ESA, and B. Whitmore) at the same scale. In the ESO Very Large Telescope image of M33, Hα is shown in red, *r*-band in orange, and *g*-band in cyan. In the Hubble Space Telescope (HST) image of NGC 4038-4039, F658N (Hα+[N II]) is shown in pink, F435W in blue, F550M in green, and F814W in red. The large HII region NGC 604 (Credit: NASA and The Hubble Heritage Team) within M33 is highlighted and shown as a composite of ten different HST filters. These two examples show two different star forming galaxy morphologies, which supports the use of an LOC model.

The individual galaxies that make up our SF composite spectra (§2) typically have morphologies suggesting multiple star-formation sites within them. In the following models (§3), we envision a galaxy containing several tens of Giant Extragalactic H II Regions (GEHR) spread out over a region up to a few kpc across. The Stromgren spheres of the different individual GEHRs may or may not overlap. This makes the choice of an LOC model ideal, since it is highly unlikely that each individual H II region within a galaxy would experience the same ionizing flux and have the same hydrogen density. Since the sources of ionizing radiation are likely distributed throughout SF galaxies rather than centrally concentrated as in AGN, we interpret the parameter describing the radial distribution of the ionized gas as actually describing the distribution of ionizing flux as one considers different locations within the same SF galaxy. Otherwise the methodology is the same as was used in Paper II for AGN.

As was the case in Paper II for AGN galaxies, the goal of this work is to find the physically meaningful parameters needed to reproduce the systematic variation of the emission-line properties of SF galaxies covering the SF wing of the BPT diagram. In Paper II we found that the positions of active galaxies along the AGN portion of the BPT diagram are primarily determined by a single parameter, the variation in the central concentration of gas clouds, but other likely parameters for the SF galaxies studied in this present paper include star formation history (SFH) and metallicity.

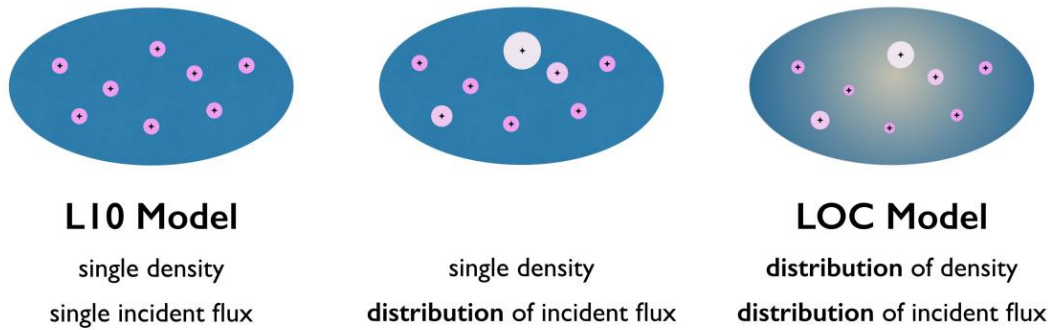

**Figure 2.** A cartoon highlighting difference between of the setups of the L10 model and an LOC model. The blue shading of the galaxy represents the density, while the pink shading within each circle represents the incident ionizing flux seen by a cloud. The circles within each galaxy represent the size of the H II region. The L10 model assumes a single density and a single incident ionizing flux value for the entire galaxy. An LOC model assumes a distribution of densities and a distribution of incident ionizing flux values are present.

In §2, we describe our method for obtaining a sample of low redshift star forming galaxies spanning a large range of ionization and the average observed properties of these galaxies. Then in §3 we present our modeling methodology and describe a large number of diagnostic diagrams that provide insight about the physical parameter responsible for the SF sequence of galaxies. In §4, we discuss the physically meaningful parameter that dictates a galaxy's position along the BPT diagram and compare our models to previous work. Finally, in §5 we summarize our conclusions and outline the needs in future modeling.

## 2. Sample Selection

For star forming galaxies with lower levels of ionization, the amount of excitation attributed to starlight becomes less clear due to possible mixing with emission line properties resulting from an active galactic nucleus. This is illustrated in Fig. 3, which shows the BPT diagram for the sample of galaxies used here. The solid and dashed green lines represent classification curves from Kewley et al. (2001) and Kauffmann et al. (2003), and respectively define the upper limit for finding SF galaxies and lower limit for finding AGN. The dotted green line extending into the upper-right corner of the diagram represents the Kauffmann et al. (2003) line that divides a class of low ionization objects (LINERs) from more classical AGN. The correct classification of AGN and SF galaxies is readily apparent at higher ionization due to the large degree of separation in [N II] λ6584/ Hα above log([O III]/Hβ]) ~ 0.3. However below this threshold, the two "branches" of AGN and SF galaxies merge to form a region containing a mix of star-forming galaxies, AGN and possible composite cases. The actual theoretical boundary between AGN and SF galaxies, and the degree of mixing between the two, is a matter of debate (Kewley et al 2001; Kauffmann et al. 2003; Kewley et al. 2006; Stasińska et al. 2006). Therefore, it is likely that previous models are limited in their interpretation of galaxies that lie inside the composite region. Furthermore, most analysis of optical-passband emission lines has been limited to classical line ratio diagrams from VO87. Additional diagrams would provide further constraints or consistency checks on models developed from these strong line techniques.

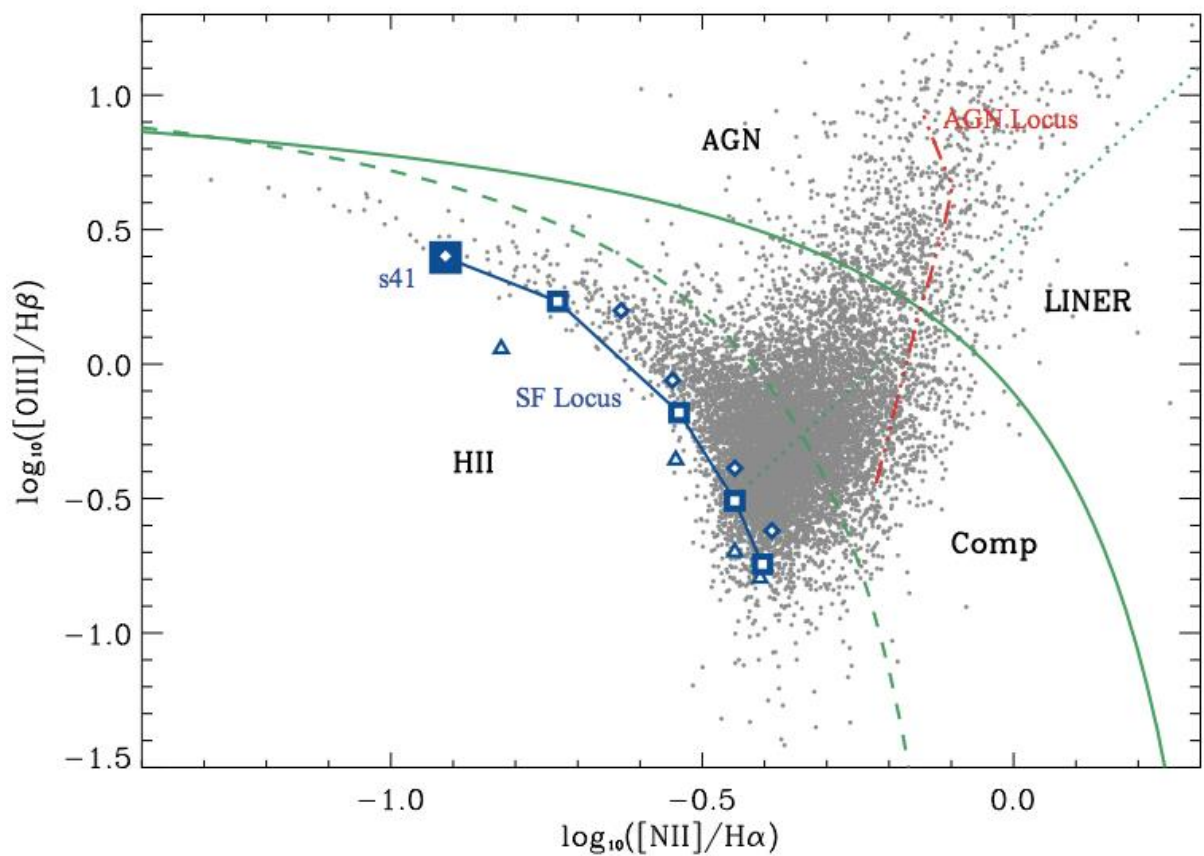

**Figure 3.** The BPT diagram with our sample of galaxies plotted as grey dots. The solid green line is the theoretical upper limit for SF galaxies presented by Kewley et al. (2001). The dashed green line is the classification curve used by Kauffmann et al. (2003) as a lower limit for finding AGN. The dotted green line shows the division of AGN and LINERs from Kauffmann et al. (2003). The solid blue curve is the SF locus and the dashed red curve is the AGN locus from our MFICA analysis. The blue squares show the dereddened ratios for the sequence of subsets along the SF locus listed in Table 2, with s41 representing the highest ionization subset. The blue triangles and diamonds are the MFICA "wing sequences" described in the text; the SF locus and wing sequences are coincident for the highest-ionization case, but not for the other subsets.

Here we present an alternative look at this issue, starting with a cleaner sample of SF galaxies identified using a new statistical technique. In Paper I, we showed that a variant of blind source separation, known as mean field independent component analysis (MFICA), successfully alleviates the restriction of mixed emission line components from AGN and starlight excitation that occurs on the BPT and similar diagrams. We refer the reader to Papers I and II for a more detailed description of the MFICA procedure, but in summary this approach has many similarities to principal component analysis (Francis et al. 1992), except that it uses a different algorithm that produces components with greater physical meaning.

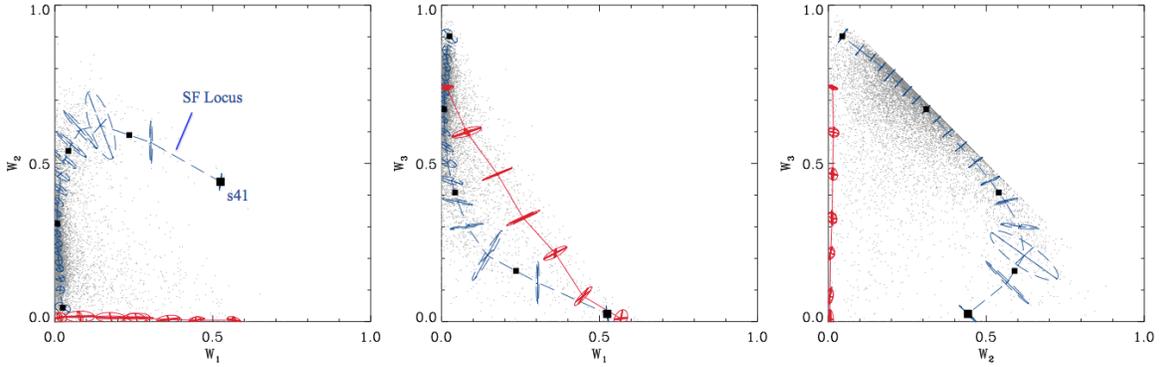

**Figure 4.** Distributions of the MFICA weights for each component. In this representation the SF locus appears as the dashed blue line, and the AGN locus is the solid red line. The ovals along each locus represent the scatter of data points still identified as AGN or pure SF. The locations of the subsets along the central SF locus described in §2 are marked by black squares with the highest-ionization SF subset s41 labeled in the far left panel.

We used the MFICA technique on small subsets of Sloan Digital Sky Survey (SDSS) emission-line galaxies to isolate a series of continuum and emission line components that combine in differing amounts to produce the observed spectra. We used a subset of 170 galaxies to generate the continuum components and 727 galaxies to generate the emission line components (Paper I). We then fitted these components to the full sample of ~$10^4$ low redshift ($0.10 < z < 0.12$) SDSS galaxies, and used the weights of each of the 5 emission-line components in each galaxy's spectrum as the coordinates of that galaxy in a 5-dimensional parameter space. The unique signatures of AGN and SF galaxies in that 5-dimensional MFICA parameter space allowed us to use an automated technique to isolate a sequence of "pure" AGN galaxies over a large range of ionization reaching well down into the "composite" region on the BPT diagram. This also allowed us to verify that our pre-selected set of SF galaxies were excited only by starlight. Our sample size also enabled many weaker lines to be accurately measured.

We found that pure SF galaxies can be reliably identified and placed in order along an "SF sequence" using only the first three MFICA emission-line components. Fig. 4, which is similar to Fig. 3 from Paper II, shows a more abstract replacement for the BPT diagram, which are two dimensional projections of the parameter space formed by the first three MFICA components, with the axes now representing the weights (relative strengths) of each component. The blue and red lines present the SF and AGN loci, respectively, with the ellipses indicating the scatter among the galaxies that we identify as clearly belonging to one or the other locus. The detailed method of generating these loci is described in Papers I and II, however at a glance one can see the main point of the diagrams: there is substantial separation between the two sequences. All but the lowest ionization point of each sequence occupies a unique region of parameter space.

We measured the intensity ratios of the galaxies falling along these sequences (§2) and placed

| Table 1. Properties of the s*ij* subsets | | | | | | | | | | | | | | | |
|---|---|---|---|---|---|---|---|---|---|---|---|---|---|---|---|
| Subset | s00 | s01 | s02 | s10 | s11 | s12 | s20 | s21 | s02 | s30 | s31 | s02 | s40 | s41 | s42 |
| $E$(B-V) | 0.69 | 0.68 | 0.67 | 0.37 | 0.44 | 0.49 | 0.28 | 0.28 | 0.30 | 0.15 | 0.20 | 0.27 | 0.14 | 0.14 | 0.14 |
| Observed $L$(Hβ)[1] | 2.7 | 2.8 | 2.8 | 2.8 | 3.5 | 3.4 | 4.3 | 7.3 | 12. | 11. | 13. | 18. | 23. | 22. | 21. |
| Observed $L$([O III] 5007)[1] | 0.48 | 0.57 | 0.72 | 0.59 | 1.2 | 1.5 | 2.0 | 4.9 | 11. | 12. | 24. | 29. | 57. | 57. | 55. |
| Observed $L_\lambda(\lambda 5007)$[2] | 0.53 | 0.53 | 0.54 | 0.46 | 0.48 | 0.50 | 0.46 | 0.57 | 0.65 | 0.55 | 0.61 | 0.65 | 0.67 | 0.70 | 0.70 |
| Dereddened $L^c$(Hβ)[1] | 29.5 | 30.1 | 28.6 | 10.2 | 15.9 | 18.9 | 11.4 | 19.4 | 35.6 | 17.8 | 26.5 | 45.4 | 37.1 | 36.4 | 34.5 |
| Dereddened $L^c$([O III] 5007)[1] | 4.71 | 5.42 | 6.88 | 2.04 | 4.93 | 7.76 | 5.00 | 12.8 | 30.9 | 20.3 | 45.3 | 71.7 | 92.0 | 90.7 | 86.6 |
| Dereddened $L^c_\lambda(\lambda 5007)$[2] | 5.27 | 5.07 | 5.14 | 1.59 | 2.06 | 2.67 | 1.17 | 1.48 | 1.81 | 0.91 | 1.18 | 1.62 | 1.08 | 1.11 | 1.10 |
| $W_\lambda$(Hβ) (Å) | 5.04 | 5.39 | 5.15 | 5.93 | 6.79 | 6.41 | 9.00 | 12.4 | 18.0 | 19.3 | 21.0 | 26.8 | 32.3 | 31.8 | 30.4 |
| $W_\lambda$([O III] 5007) (Å) | 0.91 | 1.08 | 1.34 | 1.25 | 2.24 | 2.76 | 4.14 | 8.43 | 16.2 | 22.4 | 36.9 | 43.6 | 81.4 | 80.9 | 77.7 |
| Component weights | | | | | | | | | | | | | | | |
| Component 1 | 0.00 | 0.02 | 0.05 | 0.00 | 0.01 | 0.02 | 0.00 | 0.04 | 0.10 | 0.18 | 0.24 | 0.30 | 0.53 | 0.52 | 0.52 |
| Component 2 | 0.06 | 0.04 | 0.02 | 0.32 | 0.31 | 0.29 | 0.57 | 0.54 | 0.50 | 0.74 | 0.59 | 0.43 | 0.47 | 0.44 | 0.40 |
| Component 3 | 0.93 | 0.90 | 0.87 | 0.68 | 0.67 | 0.65 | 0.42 | 0.41 | 0.39 | 0.05 | 0.16 | 0.27 | 0.00 | 0.02 | 0.06 |

[1]Luminosities are in units of $10^{40}$ erg s$^{-1}$.
[2]Continuum luminosities are in units of $10^{40}$ erg s$^{-1}$ Å$^{-1}$.

them on top of the galaxies in the BPT diagram in Fig. 3. The squares in Figs. 3 and 4 show the positions of each subsample of galaxies along the SF locus, while the dashed red line shows the AGN locus. The separation between the SF and AGN loci is well defined even at the lowest ionization. We also identify two parallel wing sequences of SF galaxies, on opposite sides of the central sequence that (as was discussed in Papers I and II) represent the scatter of true SF galaxies in directions orthogonal to the central SF locus. These wing sequences, indicated on Fig. 3 by blue diamonds and triangles, represent the data points along the edges of the ellipses shown in Fig. 4. They show that the SF locus is quite narrow in comparison to its length (as is the AGN locus). This illustrates the key point that a single physical parameter might be responsible for at least most of the variation along either the AGN or the SF locus in emission line galaxies although that single parameter might be different for the SF and AGN cases. We seek to understand that single physical parameter for the SF locus in this paper.

Composite observed emission line spectra were formed for each locus point in the following manner. We first used MFICA on each individual galaxy's spectrum to determine the continuum components (following the procedures described in Paper I) and we then subtracted the reconstructed continuum shape from the galaxy's spectrum. Next, a composite emission line spectrum was generated at each locus point by co-adding the continuum subtracted spectra in each subsample. We adopt the same nomenclature as in Paper II by naming the points along the sequence in the format s*ij*. The first index, *i*, indicates the ionization with *i* = 4 corresponding to the highest ionization starburst galaxies and *i* = 0 corresponding to the lowest ionization SF galaxies. The second index, *j*, describes the position of the sequence orthogonal to the SF locus with *j* = 1 designating the central sequence, *j* = 0 the subsets positioned closest to galaxies with AGN and *j* = 2 the subsets located on the opposite side. Table 1 lists the MFICA component weights that define each of the subsamples along the SF sequence. From the large SDSS sample, we identified for each subsample the 50 galaxies that lie closest to the defining point of that subsample in terms of their MFICA component weights, and co-added their spectra to make an average subsample spectrum. Table 1 lists for each subsample the reddening *E(B-V)* determined from the Hα/Hβ intensity ratio, the dereddened Hβ and [O III] λ5007 luminosities, denoted as $L^c$(Hβ) and $L^c$([O III] 5007), and the continuum luminosity, $L^c_\lambda(\lambda 5007)$, in addition to the component weights for each subsample. Finally, the equivalent widths for Hβ and [O III] λ5007, denoted $W_\lambda$(Hβ) and $W_\lambda$([O III] λ5007) are given. Fig. 5 displays several regions of the co-added spectra for our central sequence; the wing sequences are not displayed due to their close similarity to the central sequence.

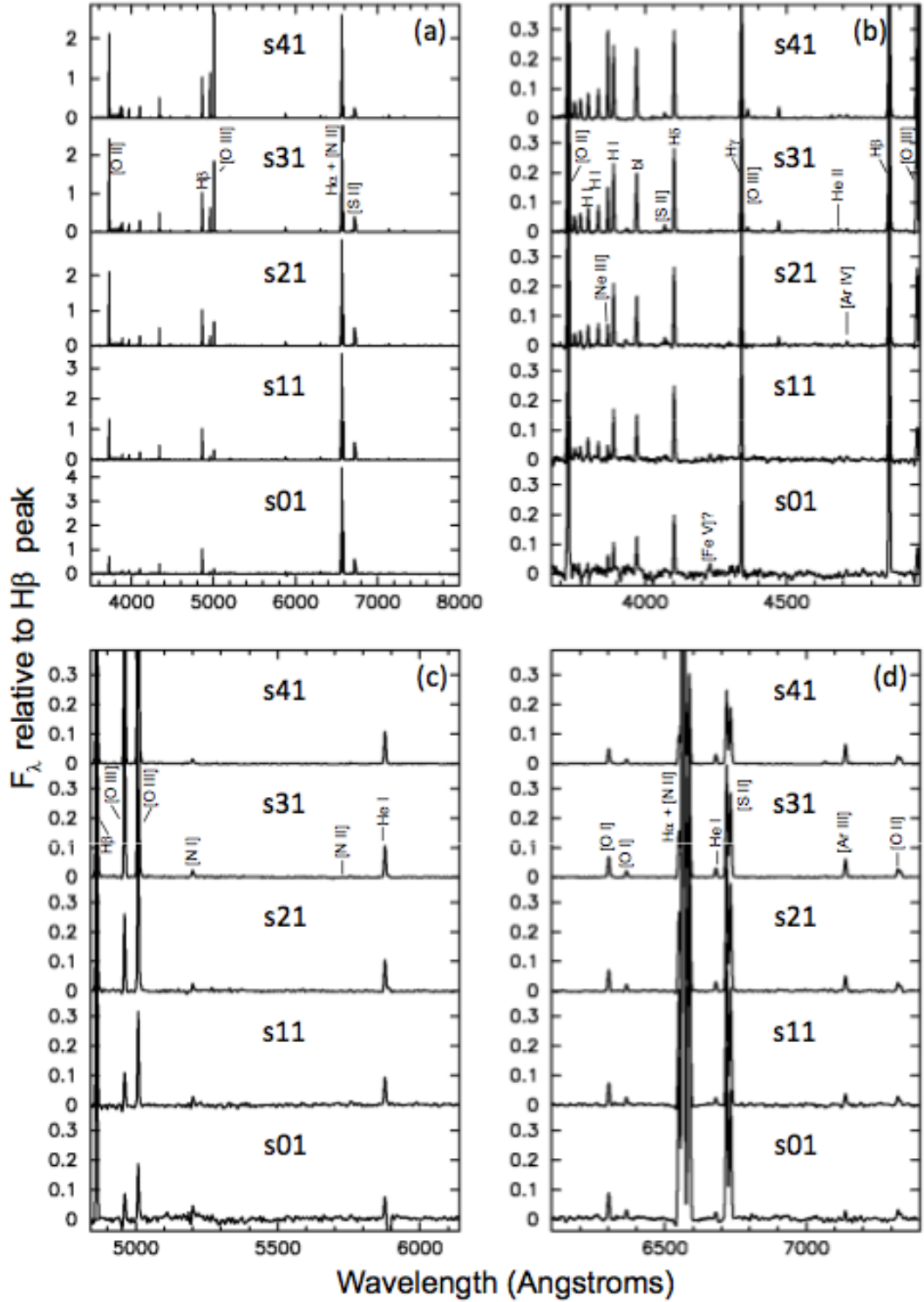

**Figure 5.** Observed, coadded spectra for the five subsamples that fall directly along the central SF locus. Panel (a) shows the full spectra. The other panels show enlargements that slightly overlap in wavelength. The emission lines listed in Table 2 are identified by their ions. "bl" indicates a blended line. All flux values are shown in units of the peak Hβ intensity in the particular spectrum.

As was done in Paper II for the AGN sample, the emission lines present in the co-added spectra of these SF galaxies were measured by subtracting a locally fitted continuum for each spectrum and then integrating the flux within each line profile. Since the earlier subtraction of the reconstructed galaxy continuum removed underlying absorption lines and other continuum structure from the spectra of the individual galaxies, the residual continuum in the coadded spectra is only a small and fairly smoothly varying residual, so a simple low-order polynomial continuum fit was adequate when actually measuring the emission lines. Table 2 lists our measured values for the observed emission line fluxes.

The classical strong emission lines are marked in Fig. 5a. They are all detected with high signal:noise in all spectra. [O II] λ3727, Hβ and [O III] λ5007 are free of significant blending, and we adopt 15 percent as a reasonable estimate of the uncertainty in our flux measurements, set more by the probability of systematic calibration errors than by the pixel-to-pixel noise. Hα is partially blended with [N II] λλ6548, 6584, and its flux was measured by integrating over wavelength between the lowest points in the blended region between it and the adjacent [N II] lines. Since Hα contributes over 2/3 of the total flux in the blend, we estimate that it is still measured to about 20 percent accuracy, while the weaker [N II] λ6584 line is probably measured to about 30 percent accuracy. The [S II] λλ6716, 6731 lines are more heavily blended, and were separated by integrating over wavelength to the lowest point in the blended region between them. While the total strength in the doublet was measured to an estimated 15 percent accuracy, we assign an uncertainty of 30 percent to the measurements of the individual lines. These estimated uncertainties will be used below to determine characteristic uncertainties in the various line intensity ratios that will be compared to our models.

Panels (b), (c) and (d) of Fig. 5 show a number of additional, weaker lines which we also measured (Table 2). We mention here only those lines that are of interest in our analysis below. We estimate typical uncertainties of 30 percent for the intermediate strength lines [Ne III] λ3869, [O I] λ6300 and [Ar III] λ7135. The He I λ5876 line is also fairly strong, but it is blended with residual Na D absorption that was not always removed very well when we subtracted off the reconstructed spectra of the underlying galaxies (see the bottom spectrum in Fig 5c). We measured the He I line by fitting its lower-wavelength half with the Hβ profile, and we adopt a factor of two uncertainty in the resulting flux.

Our large sample of pure SF galaxies (50 galaxies per subset) has also enabled measurements (or in some cases useful upper limits) for several very weak lines that are not usually detectable in lower-quality spectra, including [S II] λ4070, [O III] λ4363, He II λ4686, [Ar IV] λ4711, [N I] λ5200, [N II] λ5755 and [O II] λ7325. We adopt a factor of two uncertainty for the flux measurements for these lines.

Table 3 lists dereddened line strengths, assuming for simplicity a standard Galactic $R_V$ = 3.1 reddening curve (Cardelli, Clayton & Mathis 1989) with an *E(B-V)* that gives a dereddened *I*(Hα)/*I*(Hβ) = 2.86, appropriate for Case B recombination lines with an electron temperature $T_e$ = $10^4$ K and an electron density $n_e$ = $10^2$ cm$^{-3}$ (Osterbrock & Ferland 2006; hereafter AGN3). We compare our models below to the dereddened line strengths. The dereddening step adds considerable further uncertainty to the intensity ratios of widely separated lines, so in our

| Table 2. The measured emission line strengths for the SF locus. Measurements are relative to Hβ. | | | | | | | | | | | | | | | | |
|---|---|---|---|---|---|---|---|---|---|---|---|---|---|---|---|---|
| Ion | λ | s00 | s01 | s02 | s10 | s11 | s12 | s20 | s21 | s22 | s30 | s31 | s32 | s40 | s41 | s42 |
| [O II] | 3727 | 0.73 | 0.74 | 0.77 | 1.28 | 1.34 | 1.39 | 2.05 | 2.01 | 1.92 | 2.66 | 2.36 | 2.08 | 2.11 | 2.10 | 2.08 |
| H I | 3750 | <0.03 | <0.03 | <0.04 | 0.04 | 0.03 | 0.03 | 0.04 | 0.03 | 0.05 | 0.05 | 0.04 | 0.04 | 0.04 | 0.04 | 0.04 |
| H I | 3771 | <0.01 | <0.01 | <0.00 | 0.03 | 0.04 | 0.01 | 0.05 | 0.04 | 0.04 | 0.05 | 0.05 | 0.05 | 0.05 | 0.05 | 0.05 |
| H I | 3798 | <0.00 | <0.01 | <0.02 | 0.05 | 0.06 | 0.04 | 0.05 | 0.06 | 0.06 | 0.06 | 0.06 | 0.06 | 0.06 | 0.06 | 0.06 |
| H I | 3835 | <0.02 | <0.03 | <0.04 | 0.06 | 0.05 | 0.05 | 0.05 | 0.05 | 0.07 | 0.07 | 0.07 | 0.07 | 0.07 | 0.07 | 0.07 |
| [Ne III] | 3869 | 0.05 | 0.05 | 0.05 | 0.07 | 0.03 | 0.06 | 0.05 | 0.05 | 0.06 | 0.15 | 0.11 | 0.10 | 0.22 | 0.22 | 0.23 |
| H I | 3889 | 0.08 | 0.09 | 0.09 | 0.14 | 0.13 | 0.12 | 0.16 | 0.16 | 0.18 | 0.19 | 0.18 | 0.18 | 0.20 | 0.20 | 0.20 |
| H I | 3970 | 0.11 | 0.13 | 0.14 | 0.11 | 0.13 | 0.14 | 0.13 | 0.14 | 0.16 | 0.19 | 0.18 | 0.18 | 0.22 | 0.22 | 0.23 |
| [S II] | 4070 | <0.05 | <0.04 | <0.01 | <0.01 | <0.02 | 0.03 | 0.03 | 0.03 | 0.02 | 0.03 | 0.02 | 0.02 | 0.02 | 0.02 | 0.02 |
| H I | 4102 | 0.18 | 0.18 | 0.16 | 0.20 | 0.21 | 0.21 | 0.21 | 0.21 | 0.23 | 0.24 | 0.23 | 0.24 | 0.25 | 0.25 | 0.25 |
| [Fe V]? | 4229 | 0.04 | 0.05 | 0.06 | 0.03 | 0.02 | <0.01 | <0.01 | <0.01 | <0.00 | <0.01 | <0.00 | <0.01 | <0.00 | <0.00 | <0.00 |
| H I | 4340 | 0.34 | 0.34 | 0.34 | 0.40 | 0.40 | 0.40 | 0.40 | 0.42 | 0.42 | 0.43 | 0.43 | 0.43 | 0.45 | 0.45 | 0.45 |
| [O III] | 4363 | <0.00 | <0.00 | <0.00 | <0.00 | <0.00 | <0.00 | <0.00 | <0.00 | <0.01 | 0.01 | 0.01 | 0.01 | 0.03 | 0.03 | 0.03 |
| He I | 4471 | <0.01 | <0.00 | <0.00 | <0.02 | <0.00 | <0.00 | 0.03 | 0.02 | 0.02 | 0.03 | 0.03 | 0.03 | 0.03 | 0.03 | 0.03 |
| He II | 4686 | <0.00 | <0.00 | <0.01 | <0.01 | <0.01 | <0.00 | <0.00 | <0.00 | <0.01 | 0.00 | 0.01 | 0.01 | 0.01 | 0.01 | 0.01 |
| [Ar IV] | 4711 | <0.01 | <0.01 | <0.01 | <0.03 | <0.02 | <0.02 | 0.02 | 0.01 | 0.01 | 0.01 | 0.01 | 0.01 | 0.01 | 0.01 | 0.01 |
| H I | 4861 | 1.00 | 1.00 | 1.00 | 1.00 | 1.00 | 1.00 | 1.00 | 1.00 | 1.00 | 1.00 | 1.00 | 1.00 | 1.00 | 1.00 | 1.00 |
| [O III] | 4959 | 0.07 | 0.08 | 0.10 | 0.09 | 0.12 | 0.15 | 0.18 | 0.25 | 0.30 | 0.68 | 0.58 | 0.55 | 1.05 | 1.06 | 1.08 |
| [O III] | 5007 | 0.18 | 0.20 | 0.26 | 0.21 | 0.33 | 0.43 | 0.46 | 0.68 | 0.90 | 1.16 | 1.76 | 1.63 | 2.52 | 2.54 | 2.56 |
| [N I] | 5200 | 0.04 | 0.05 | 0.04 | 0.02 | 0.03 | 0.04 | 0.03 | 0.02 | 0.03 | 0.03 | 0.03 | 0.03 | 0.02 | 0.02 | 0.02 |
| [N II] | 5755 | <0.01 | <0.02 | <0.01 | <0.01 | <0.02 | <0.01 | <0.01 | <0.00 | <0.01 | 0.00 | 0.00 | 0.01 | 0.00 | 0.00 | 0.00 |
| He I | 5876 | 0.09 | 0.09 | 0.09 | 0.09 | 0.11 | 0.12 | 0.12 | 0.12 | 0.13 | 0.12 | 0.13 | 0.14 | 0.13 | 0.13 | 0.13 |
| [O I] | 6300 | 0.10 | 0.10 | 0.11 | 0.07 | 0.08 | 0.11 | 0.09 | 0.08 | 0.09 | 0.09 | 0.09 | 0.09 | 0.06 | 0.06 | 0.06 |
| [O I] | 6363 | 0.03 | 0.03 | 0.05 | 0.02 | 0.03 | 0.03 | 0.03 | 0.02 | 0.03 | 0.03 | 0.03 | 0.03 | 0.02 | 0.02 | 0.02 |
| [N II] | 6548 | 0.70 | 0.69 | 0.71 | 0.43 | 0.48 | 0.51 | 0.32 | 0.33 | 0.34 | 0.14 | 0.20 | 0.28 | 0.13 | 0.13 | 0.13 |
| H I | 6563 | 5.86 | 5.79 | 5.74 | 4.21 | 4.49 | 4.77 | 3.82 | 3.83 | 3.92 | 3.34 | 3.51 | 3.78 | 3.31 | 3.30 | 3.29 |
| [N II] | 6584 | 2.32 | 2.30 | 2.36 | 1.51 | 1.61 | 1.71 | 1.10 | 1.11 | 1.11 | 0.50 | 0.66 | 0.89 | 0.41 | 0.41 | 0.40 |
| He I | 6678 | 0.02 | 0.02 | 0.03 | 0.03 | 0.03 | 0.04 | 0.03 | 0.04 | 0.04 | 0.04 | 0.04 | 0.04 | 0.04 | 0.04 | 0.04 |
| [S II] | 6716 | 0.77 | 0.78 | 0.79 | 0.66 | 0.72 | 0.78 | 0.71 | 0.64 | 0.59 | 0.51 | 0.50 | 0.49 | 0.33 | 0.33 | 0.32 |
| [S II] | 6731 | 0.59 | 0.59 | 0.59 | 0.50 | 0.53 | 0.59 | 0.52 | 0.49 | 0.45 | 0.38 | 0.38 | 0.39 | 0.26 | 0.26 | 0.25 |
| [Ar III] | 7135 | 0.02 | 0.03 | 0.04 | 0.04 | 0.05 | 0.05 | 0.06 | 0.06 | 0.07 | 0.07 | 0.08 | 0.09 | 0.09 | 0.09 | 0.09 |
| [O II] | 7325 | 0.07 | 0.07 | 0.08 | 0.02 | 0.06 | 0.06 | 0.06 | 0.05 | 0.06 | 0.06 | 0.06 | 0.07 | 0.06 | 0.06 | 0.06 |

**Note.** Values shown as 0.00 indicate lines which are detected but whose measured line strengths round down to 0.00. Values shown as <0.00 indicate lines with upper limits that round down to 0.00.

**Table 3.** The dereddened emission line strengths for the SF locus. Measurements are relative to Hβ.

| Ion | λ | s00 | s01 | s02 | s10 | s11 | s12 | s20 | s21 | s22 | s30 | s31 | s32 | s40 | s41 | s42 |
|---|---|---|---|---|---|---|---|---|---|---|---|---|---|---|---|---|
| [O II] | 3727 | 1.59 | 1.59 | 1.63 | 1.94 | 2.17 | 2.41 | 2.80 | 2.75 | 2.70 | 3.15 | 2.94 | 2.81 | 2.46 | 2.45 | 2.42 |
| H I | 3750 | <0.06 | <0.07 | <0.08 | 0.06 | 0.05 | 0.05 | 0.05 | 0.04 | 0.06 | 0.05 | 0.05 | 0.06 | 0.05 | 0.05 | 0.05 |
| H I | 3771 | <0.03 | <0.02 | <0.01 | 0.05 | 0.06 | 0.02 | 0.07 | 0.05 | 0.06 | 0.06 | 0.06 | 0.07 | 0.06 | 0.06 | 0.06 |
| H I | 3798 | <0.01 | <0.03 | <0.04 | 0.07 | 0.10 | 0.06 | 0.07 | 0.07 | 0.08 | 0.08 | 0.08 | 0.08 | 0.07 | 0.07 | 0.07 |
| H I | 3835 | <0.04 | <0.06 | <0.08 | 0.09 | 0.07 | 0.08 | 0.07 | 0.07 | 0.09 | 0.08 | 0.08 | 0.09 | 0.08 | 0.08 | 0.08 |
| [Ne III] | 3869 | 0.11 | 0.10 | 0.09 | 0.10 | 0.05 | 0.09 | 0.07 | 0.06 | 0.08 | 0.17 | 0.14 | 0.13 | 0.25 | 0.26 | 0.26 |
| H I | 3889 | 0.17 | 0.18 | 0.17 | 0.20 | 0.20 | 0.20 | 0.21 | 0.21 | 0.24 | 0.22 | 0.22 | 0.24 | 0.23 | 0.23 | 0.23 |
| H I | 3970 | 0.20 | 0.24 | 0.26 | 0.15 | 0.19 | 0.21 | 0.17 | 0.18 | 0.20 | 0.22 | 0.22 | 0.23 | 0.25 | 0.25 | 0.26 |
| [S II] | 4070 | <0.08 | <0.06 | <0.02 | <0.02 | <0.02 | 0.04 | 0.04 | 0.04 | 0.03 | 0.04 | 0.03 | 0.03 | 0.02 | 0.02 | 0.02 |
| H I | 4102 | 0.30 | 0.30 | 0.28 | 0.27 | 0.29 | 0.31 | 0.27 | 0.26 | 0.29 | 0.27 | 0.27 | 0.30 | 0.28 | 0.28 | 0.28 |
| [Fe V]? | 4229 | 0.07 | 0.07 | 0.10 | 0.04 | 0.02 | 0.01 | <0.01 | <0.01 | <0.00 | <0.01 | <0.00 | <0.01 | <0.00 | <0.00 | <0.00 |
| H I | 4340 | 0.49 | 0.50 | 0.49 | 0.50 | 0.51 | 0.52 | 0.46 | 0.49 | 0.50 | 0.47 | 0.48 | 0.50 | 0.49 | 0.48 | 0.48 |
| [O III] | 4363 | <0.00 | <0.01 | <0.00 | <0.00 | <0.00 | <0.00 | <0.00 | <0.00 | <0.01 | 0.01 | 0.01 | 0.01 | 0.03 | 0.03 | 0.03 |
| He I | 4471 | <0.01 | <0.01 | <0.00 | <0.02 | <0.00 | <0.00 | 0.03 | 0.02 | 0.03 | 0.03 | 0.03 | 0.03 | 0.03 | 0.03 | 0.03 |
| He II | 4686 | <0.00 | <0.01 | <0.01 | <0.01 | <0.01 | <0.00 | <0.00 | <0.00 | <0.01 | 0.00 | 0.01 | 0.01 | 0.01 | 0.01 | 0.01 |
| [Ar IV] | 4711 | <0.01 | <0.01 | <0.01 | <0.03 | <0.02 | <0.02 | 0.02 | 0.01 | 0.01 | 0.01 | 0.01 | 0.01 | 0.01 | 0.01 | 0.01 |
| H I | 4861 | 1.00 | 1.00 | 1.00 | 1.00 | 1.00 | 1.00 | 1.00 | 1.00 | 1.00 | 1.00 | 1.00 | 1.00 | 1.00 | 1.00 | 1.00 |
| [O III] | 4959 | 0.07 | 0.08 | 0.10 | 0.09 | 0.11 | 0.15 | 0.17 | 0.24 | 0.29 | 0.67 | 0.57 | 0.54 | 1.04 | 1.05 | 1.07 |
| [O III] | 5007 | 0.16 | 0.18 | 0.24 | 0.20 | 0.31 | 0.41 | 0.44 | 0.66 | 0.87 | 1.14 | 1.71 | 1.58 | 2.48 | 2.49 | 2.52 |
| [N I] | 5200 | 0.03 | 0.04 | 0.04 | 0.02 | 0.03 | 0.03 | 0.03 | 0.02 | 0.03 | 0.03 | 0.03 | 0.03 | 0.02 | 0.02 | 0.02 |
| [N II] | 5755 | <0.01 | <0.01 | <0.01 | <0.01 | <0.01 | <0.01 | <0.00 | <0.00 | <0.01 | 0.00 | 0.00 | 0.01 | 0.00 | 0.00 | 0.00 |
| He I | 5876 | 0.06 | 0.05 | 0.05 | 0.07 | 0.08 | 0.09 | 0.10 | 0.10 | 0.10 | 0.11 | 0.11 | 0.11 | 0.12 | 0.12 | 0.12 |
| [O I] | 6300 | 0.05 | 0.05 | 0.06 | 0.05 | 0.06 | 0.07 | 0.07 | 0.06 | 0.07 | 0.08 | 0.07 | 0.07 | 0.06 | 0.06 | 0.06 |
| [O I] | 6363 | 0.01 | 0.02 | 0.03 | 0.01 | 0.02 | 0.02 | 0.02 | 0.02 | 0.02 | 0.03 | 0.02 | 0.02 | 0.02 | 0.02 | 0.02 |
| [N II] | 6548 | 0.34 | 0.34 | 0.36 | 0.29 | 0.31 | 0.31 | 0.24 | 0.25 | 0.25 | 0.12 | 0.16 | 0.21 | 0.11 | 0.11 | 0.11 |
| H I | 6563 | 2.86 | 2.86 | 2.86 | 2.86 | 2.86 | 2.86 | 2.86 | 2.86 | 2.86 | 2.86 | 2.86 | 2.86 | 2.86 | 2.86 | 2.86 |
| [N II] | 6584 | 1.12 | 1.13 | 1.17 | 1.02 | 1.02 | 1.02 | 0.82 | 0.83 | 0.81 | 0.43 | 0.53 | 0.67 | 0.35 | 0.35 | 0.35 |
| He I | 6678 | 0.01 | 0.01 | 0.02 | 0.02 | 0.02 | 0.02 | 0.02 | 0.03 | 0.03 | 0.03 | 0.03 | 0.03 | 0.03 | 0.03 | 0.03 |
| [S II] | 6716 | 0.36 | 0.37 | 0.37 | 0.44 | 0.44 | 0.45 | 0.52 | 0.47 | 0.42 | 0.44 | 0.40 | 0.36 | 0.28 | 0.28 | 0.28 |
| [S II] | 6731 | 0.27 | 0.27 | 0.28 | 0.33 | 0.33 | 0.34 | 0.38 | 0.35 | 0.32 | 0.33 | 0.31 | 0.29 | 0.22 | 0.22 | 0.22 |
| [Ar III] | 7135 | 0.01 | 0.01 | 0.02 | 0.03 | 0.03 | 0.03 | 0.04 | 0.04 | 0.05 | 0.06 | 0.06 | 0.06 | 0.07 | 0.07 | 0.07 |
| [O II] | 7325 | 0.03 | 0.03 | 0.03 | 0.01 | 0.03 | 0.03 | 0.04 | 0.04 | 0.04 | 0.05 | 0.05 | 0.05 | 0.05 | 0.05 | 0.05 |

**Note.** Values shown as 0.00 indicate lines which are detected but whose measured line strengths round down to 0.00. Values shown as <0.00 indicate lines with upper limits that round down to 0.00.

analysis below we have assigned an additional 30 per cent uncertainty to the ratios of lines separated by more than 1000Å.

Our sample of star-forming galaxies have two distinct morphologies. The highly ionized galaxies ([O III]/ Hβ > 1) are commonly classified (~40%) as mergers in the Galaxy Zoo catalog (Lintott et al. 2011). While the weakly ionized galaxies ([O III]/ Hβ < 1) are identified as either spiral or elliptical by Galaxy Zoo volunteers, a color-magnitude diagram (*g-r* vs. *r*) shows that all of our pure star-forming galaxies are well within the bounds of the "blue cloud" and do not fall along the "red sequence." Furthermore, it is only the brighter half of these galaxies that are tagged as having spiral arms; the fainter galaxies are tagged as being elliptical. Thus, we believe that the majority of the weakly ionized star-forming galaxies are spiral galaxies and that for some the spiral arms are simply not bright enough to be detectable in SDSS/Galaxy Zoo images. In fact, reclassification of SDSS ellipticals is possible in cases where more sensitive imaging with SDSS Stripe 82 is available (Schawinski et al. 2010). These morphologies are consistent with those of local star forming galaxies (Fig. 1) justifying the use of a LOC model.

A key element of our investigation concerns the importance of the spatial relationship between the ionizing stars in galaxies and the gas producing the emission-line spectra. It is thus essential that the SDSS galaxy spectra we use probe spatial scales of several kpc and include the light from a significant fraction of the star-formation within the galaxies. This is the reason that a galaxy sample with redshifts $0.10 < z < 0.12$, where the SDSS fibre-diameter of 3" corresponds to a spatial scale of 6 kpc, is selected. Kewley et al. (2005) discuss the importance of obtaining spectra including a significant fraction of the galaxy light in order to derive physical measurements, such as metallicity, that apply to the galaxy as a whole. Brinchmann et al. (2004) used more complex aperture correction method where they calculated likelihood distribution of the star formation rate for a given set of spectra with global and fiber g−r and r−i colors and then tested this method using galaxies within the SDSS with different fiber covering fractions. Following a careful analysis, both studies conclude that it is necessary to include a minimum of 20 per cent of the galaxy light and that to satisfy that condition galaxies with SDSS spectra should lie at redshifts $z > 0.04$. With a minimum redshift of more than twice the Kewley et al. (2005) threshold, our sample comfortably satisfies the constraint and for a substantial fraction of the galaxies 40 per cent of the galaxy light is included in the SDSS spectra. About 7 per cent of our galaxies (mostly at low ionization) have covering fractions below 20 per cent in the r-band; the possible effects of this on our conclusions will be discussed later in §4.4.

## 3. LOC Modeling

The LOC model is based on a grid of individual photoionization models of gas at different densities and different distances from the source. After determining the spectral energy distribution (SED) and chemical abundances, the main free parameters correspond to the distribution of the incident ionizing flux striking the clouds and the density distribution of those clouds.

We used version c13.02 of the plasma simulation code CLOUDY to compute our grid of photoionization models. As in Paper II, and Ferguson et al. (1997), we assumed there is no attenuation of the incident continuum by other ISM clouds. Our solar abundance set comes from Grevesse et al. (2010). Plasmas with $T_e < 4000$ K or $T_e > 10^5$ K do not contribute significantly to optimal emission, so we use these conditions as limits on the Cloudy models used in our simulations.

We ran grids of simulations varying the ionizing flux incident on each cloud, $\phi_H$, between $\phi_H = 10^{6.9}$ cm$^{-2}$ s$^{-1}$ and $10^{20.9}$ cm$^{-2}$ s$^{-1}$ and varying the total hydrogen density, $n_H$, between $10^0$ cm$^{-3}$ and $10^{10}$ cm$^{-3}$. The resolution of our grid resulted in 7171 total cloud models evenly distributed in the log of our $\phi_H$, $n_H$ parameter space. Each simulation assumed constant total hydrogen density, however the molecular and electron density were allowed to change with depth. The galaxies in our sample indicate multiple sites of star formation. We use 30 Doradus in the Large Magellanic Cloud as a well-studied local example of a large star-forming region. The total ionizing luminosity within the galaxy is simply,

$$L_{ion} = nL_{30dor} \quad \text{erg s}^{-1} \qquad (1)$$

where $n$ is the number of giant H II regions within the galaxy and $L_{30dor}$ is the ionizing luminosity within 30 Doradus. Taking $L_{30dor} = 10^{40.9}$ erg s$^{-1}$ (Pellegrini et al. 2011), and as a typical value $L_{ion} = 10^{42.4}$ ergs s$^{-1}$, we find that $n \sim 30$ which we find reasonable given the average mass of our SF galaxy sample, $2.5 \times 10^{10}$ M_solar. This means that, in principle, a starburst galaxy could contain up to 30 very large star-forming regions similar to 30 Doradus, or alternatively a larger number of smaller H II regions. We initially describe results using dust-free models; we will discuss simulations with dust in §3.5.

In the LOC model, the emission we observe is strongly governed by selection effects. Optimal emission line production occurs in different conditions for different emission lines. Since we only see the *cumulative* emission from many clouds each with their own local physical conditions, most of the emission in any particular line will come from the clouds for which the physical conditions are such that they emit that line most efficiently. The resulting spectrum is largely determined by the distributions of incident ionizing flux and gas densities of the clouds that produce the emission lines. We assumed that each galaxy contains a number of separate ionizing sources (e.g. star-forming regions) scattered throughout the galaxy so that in the vicinity of each such continuum source the ionizing flux is $\phi_H \propto r^{-2}$, where $r$ is the distance from the H II complex that is ionizing a particular bit of gas. Since we can no longer use $r$ to define the distance from a central ionizing continuum source as was the case in the AGN which were the subject of our previous papers (Paper II; Ferguson et al. 1997; Baldwin et al. 1995), our model is instead defined in part by the distribution of the $\phi_H$ incident on the individual ionized clouds, which we parameterize as $f(\phi_H) \propto \phi_H^\alpha$. More positive values of $\alpha$ weight regions of the LOC plane with higher ionizing flux more heavily, while smaller values of $\alpha$ weight regions that see lower ionizing flux. The other defining parameter is the distribution of gas densities among the individual clouds, for which we retain the description that we used in our earlier papers, as a power law $g(n) \propto n_H^\beta$. Here the power-law indices $\alpha$ and $\beta$ are free parameters. The total line luminosity of the whole galaxy is then the sum of the spectra of the individual star-forming regions, and follows the relation

$$L_{line} \propto \iint F(\varphi_H, n_H,)\varphi_H^\alpha n_H^\beta dn_H d\varphi_H \qquad (2)$$

where $F(\phi_H, n_H)$ is the flux of the line emitted by a particular cloud and $\phi_H{}^\alpha n_H{}^\beta \propto \psi(\phi_H, n_H)$ is the spatial distribution function. We chose the same density integration limits as in Paper II since the forbidden lines emitted by gas with $\log(n_H) > 8$ are collisionally quenched, and gas with $\log(n_H) < 2$ does not optimally emit many emission lines. The upper integration limit for the ionizing flux was arbitrarily chosen to be $\phi = 10^{16.9}$ cm$^{-2}$ s$^{-1}$. This is similar to the flux value that corresponds to the inner radial limit used for our AGN models in Paper II, and means that our LOC integrations include the peak emission from most of the higher ionization lines. It corresponds to a radial distance of about 0.02 pc from an $L_{30Dor}$ continuum source, which is much smaller than the likely size of an O-B star cluster and hence is in a region where the mean intensity of the ionizing radiation field will have leveled off. The lower flux integration limit is $\phi_H = 10^{8.9}$ cm$^{-2}$ s$^{-1}$, corresponding to a radial distance $r = 10^{20.8}$ cm = 204 pc from individual $L_{30dor}$ star-forming complexes, which is intended to represent a typical maximum distance within a galaxy from any of the individual continuum sources. This value of $\phi_H$ also corresponds to a distance $r = 10^{22}$ cm from the centroid of the summed emission from all of the SF regions within a galaxy, which is the point at which any outlying gas would start to fall outside the entrance aperture used for the SDSS spectroscopy. We experimented with changing the integration limits outside this range but changing the $\phi_H$ distribution parameter $\alpha$ and the cloud density distribution parameter $\beta$ had more noticeable effects.

With these basic assumptions, the predicted emission can constrain the properties of these galaxies by utilizing line ratio diagnostics in the same manner as VO87, and many other subsequent papers. We will first focus on evaluating the SED and metallicity before presenting classical line ratios that have been empirically determined to constrain the excitation mechanism in emission line galaxies, along with other diagrams that serve as diagnostics for other physical properties.

### 3.1 Spectral Energy Distribution

The spectral energy distribution (SED) used by our CLOUDY models was computed using the evolutionary synthesis code Starburst99 (Leitherer et al. 1999) to generate synthetic starburst spectra. We explored the Padova track evolutionary sequence with AGB stars (Bressan et al. 1993) and the Geneva track with "high" mass loss (Meynet et al. 1994), which are the default settings for the Starburst99 online simulation interface, to follow suit with other studies that have also adopted the Padova AGB evolutionary tracks (e.g. Abel & Satyapal 2008, Meléndez et al. 2014) and the Geneva "high" mass loss (e.g. L10) evolutionary tracks for their baseline models. At the time our calculations were carried out, these were the most up-to-date evolutionary tracks available for use with Starburst99 code; however, other evolutionary tracks were added after we completed our analysis. We briefly explore the impact of these tracks in §4.2.

For each track, we included the Pauldrach / Hillier model atmospheres (Pauldrach et al. 2001; Hillier & Miller 1998) for all of the SEDs. We assumed a Kroupa broken power law initial mass function (IMF; Kroupa 2001) with mass intervals of 0.1 M_solar to 0.5 M_solar and 0.5 M_solar to 100 M_solar.

This leaves two additional properties still needed to determine the SED: star formation history (SFH) and stellar (as opposed to gas) metallicity. Here, we investigate the effects on our LOC models as we vary these parameters. We used two star formation histories: instantaneous and continuous. Instantaneous starbursts assume that a population of massive stars formed with a single burst of star formation, while continuous starbursts continue to evolve until there is balance between newly born stars and the death of older stars. Our instantaneous starbursts assumed a fixed mass of $10^6$ M_solar, while our continuous starbursts assumed a star formation rate of 1 M_solar yr$^{-1}$, both of which are the default parameters for a Starburst99 simulation. For simplicity, we only considered Z_solar and 0.4Z_solar compositions. Our preliminary results showed that many of the resulting SEDs lacked enough hard photons to excite higher ionization lines. This result would only be compounded by higher metallicity spectra since the photoelectric opacity of the atmosphere is larger and thus absorbs more high-energy photons (Snijders et al. 2007). Similarly, we only consider starburst ages younger than 5 Myr. Older populations with an instantaneous SFH undergo a considerable drop-off in FUV intensity and populations with a continuous SFH have reached steady state, hence, the spectrum no longer evolves.

Fig. 6 displays the SEDs for our continuous (left panels) and instantaneous (right panels) starbursts as a function of age for solar and subsolar compositions, indicated in the upper right corner of the left panels, of populations following two different evolutionary tracks. For a stellar population that has only undergone a single burst of star formation at 0 Myr, the hardness of the SED at high energies begins to decrease as a function of age due to hot O spectral type and B spectral type stars evolving off the main sequence until the Wolf-Rayet (WR) stage significantly increases the far ultraviolet (FUV) intensity.

The SEDs shown in Fig. 6 have a wide range of hardness, and our LOC models quantify which of them carry sufficient numbers of energetic photons to produce the observed emission-line intensity ratios involving highly-ionized species in the numerator, such as He II λ4686 / Hβ, [Ar IV] λ4711 / [Ar III] λ7135, and He II λ4686 / He I λ5876. We mainly relied on the He II λ4686 / Hβ ratio for this test, since it has long been used as an indicator of the general shape of the SED (e.g. Stoy 1933; Binette et al. 1996). The oldest Padova AGB track starbursts (5 Myr) in both the continuous and instantaneous SFH models have a large amount of photons with $h\nu > 54$ eV while not having many photons around 13.6 eV, and our LOC integrations show that they produce a high He II λ4686 / Hβ ratio consistent with observations (see §3.4.2). Some previous work has preferred the spectrum generated by starbursts at a late age with a continuous SFH as opposed to those with an instantaneous SFH (Fernandes et al. 2003; L10), but other studies (Shirazi and Brinchmann 2012) found that a 3-5 Myr starburst with an instantaneous SFH is required to reproduce a large He II / Hβ ratio. Our results are consistent with *both* of these findings: a 5 Myr-old starburst with either an instantaneous *or* continuous SFH is sufficient to reproduce higher ionization emission lines. The uniting factor between these SEDs comes from the hard EUV flux due to the presence of WR stars. While both SEDs provided satisfactory fits to our higher ionization emission line ratios, the continuous SFH gave a marginally better fit. Similarly the effects of changing the stellar metallicity were marginal. Our LOC integrations also show that the younger SEDs shown in Fig. 6 do not satisfactorily reproduce the observed He II λ4686 / Hβ ratios. Thus, to simplify the subsequent analysis, for the rest of the paper we use only a 5 Myr solar metallicity Padova AGB track star cluster with a continuous SFH. However, we note that the results given below are relatively insensitive to the SFH and stellar metallicity.

## 3.2 Metallicity Optimization

After determining the best fitting SED, we turned to optimizing the gas metallicity. Emission line ratios involving nitrogen are often good determinants of metallicity due to secondary nucleosynthesis. Evidence of secondary nucleosynthesis is found in star forming galaxies (Kobulnicky & Zaritsky 1999; Liang et al. 2006) with log(N/O) up to 4 times solar values (Storchi-Bergman, Calzetti & Kinney 1994). This process creates a [N/H] $\propto Z^2$ relationship while all other metals scale linearly with metallicity (Hamann & Ferland 1999).

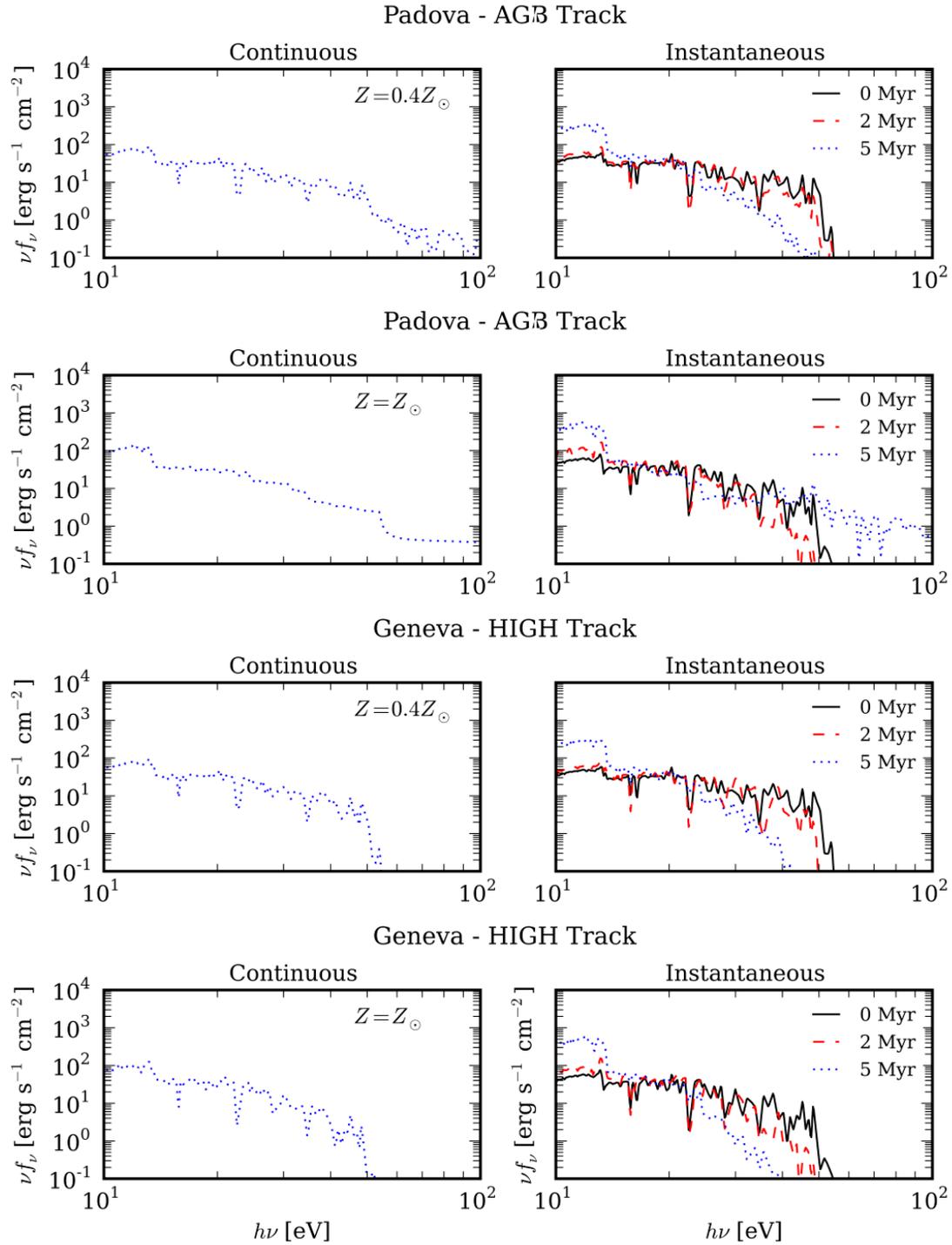

**Figure 6.** The SEDs for continuous (left panels) and instantaneous (right panels) SFHs as a function of age for clusters with solar and subsolar metallicities and two different evolutionary tracks. As the continuous starburst ages, the birth and death of stars eventually comes into equilibrium and the SED ceases to evolve. In contrast, instantaneous starbursts undergo a single period of star formation and therefore the SED continues to change as a function of age.

| Table 4. Final abundance set by number, relative to hydrogen. ||
|---|---|
| H  | 0.00    |
| He | -1.07   |
| Li | -11.17  |
| Be | -10.84  |
| B  | -9.52   |
| C  | -3.79   |
| N  | -4.61   |
| O  | -3.53   |
| F  | -7.66   |
| Ne | -4.29   |
| Na | -5.98   |
| Mg | -4.62   |
| Al | -5.77   |
| Si | -4.71   |
| P  | -6.81   |
| S  | -5.10   |
| Cl | -6.72   |
| Ar | -5.82   |
| K  | -7.19   |
| Ca | -5.88   |
| Sc | -9.07   |
| Ti | -7.27   |
| V  | -8.29   |
| Cr | -6.58   |
| Mn | -6.79   |
| Fe | -4.72   |
| Co | -7.23   |
| Ni | -6.00   |
| Cu | -8.03   |
| Zn | -7.66   |

We found that the [N II] $\lambda6584$/ [O II] $\lambda3727$ ratio, a common metallicity indicator (Groves, Heckman & Kauffmann 2006), was overpredicted by our solar abundance set. We adjusted our abundances according to the scaling relation given above in order to satisfactorily fit this ratio. We iterated over a number of subsolar metallicities[1] to find that the SF subsets are best fit by a 0.6 Z_solar metallicity. Our final abundance set by number, relative to hydrogen, in the log is given in Table 4. We will explore models with variable metallicity in §4.3.

**3.3 The LOC Plane**

Our grid of cloud models cover the two-dimensional parameter space log($n_H$) vs. log($\phi_H$), which we call the "LOC plane". Comparison of the properties of different clouds according to their positions on the LOC plane gives considerable insight into the factors controlling the final integrated spectrum emitted by our LOC model. Fig. 7 shows the LOC plane with contours of the electron temperature at the illuminated face of each cloud and also lines of constant ionization parameter. The dashed red lines display $T_e$ in 0.2 dex intervals while the solid black lines give $T_e$ in 1.0 dex intervals. The dotted blue lines give the $U = -3.0$, $U = 0.0$, $U = 3.0$ values along the LOC plane, which are even spaced with a slope given by $(n_H d\phi_H)/(\phi_H dn_H) = -1$. While there are a small number of models with $T_e > 10^5$ K or $T_e < 4000$ K, these clouds have very little effect on the overall spectrum. Fig. 8 shows equivalent width contours on the LOC plane for 20 different emission lines, all relative to the continuum at 4860 Å. The cutoff equivalent width corresponds to 1 Å while the black squares represent the peak value within each subplot. Although we computed models all the way down to $\phi_H = 10^{6.9}$ cm$^{-2}$ s$^{-1}$ which would correspond to a radial distance of about 2.0 kpc from a 30 Dor-like giant H II region, the peak emissivities for even the lowest ionization observed lines come from gas with $\phi_H$ corresponding to radial distances about 30 times smaller than this and hence well within our 200 pc integration limits. The typical spacing between the individual H II complexes in the observed galaxies, together with a plausible range in their ionizing luminosities, should therefore lead to enough low-ionization gas to produce the observed spectra. We will explore in greater detail the trends present across the LOC plane as function of SFH, gas metallicity, and grain content in a future paper (Richardson & Meskhidze 2016).

---

[1] Here we adopted the common usage of the term "metallicity" as the metals abundance relative to hydrogen, by number, normalized to solar values. However, we note that metallicity is more properly defined as a mass fraction.

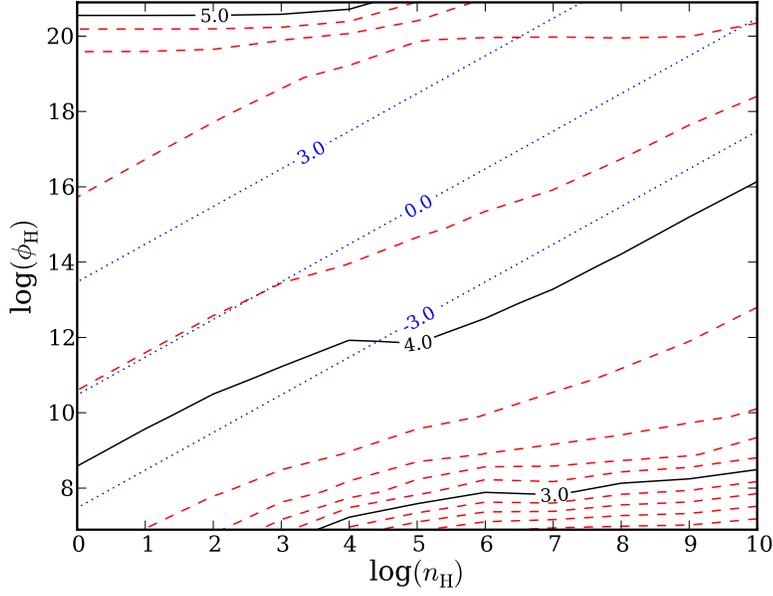

**Figure 7.** The ionization parameter and electron temperature at the illuminated face of each cloud. The ionization parameter is given by the blue contours but only for an evenly spaced representative sample across the LOC plane. The electron temperature is given in 1.0 dex (black lines) and 0.2 dex (red lines) increments.

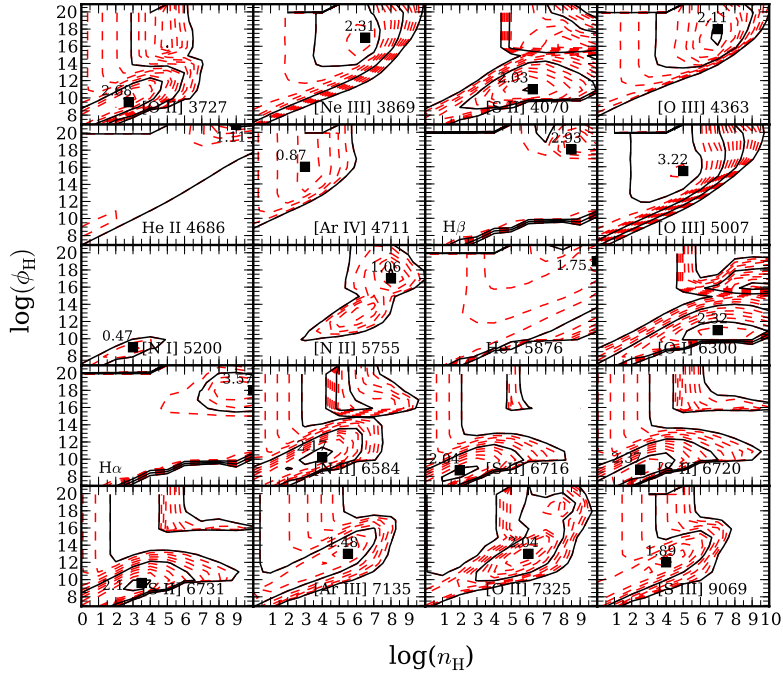

**Figure 8.** Equivalent width contours for our best model as function ionizing flux, $\varphi_H$, and hydrogen density, $n_H$. The equivalent width is expressed relative to the continuum at 4860 Å. Each labeled black square represents the maximum equivalent width in the log. Each black contour represents 1.0 dex increments while each red contour represents 0.2 dex increments.

## 3.4 Emission Line Intensity Ratio Diagrams

Using the SEDs given in §3.1 we ran LOC integrations (following the same procedures as in Paper II) to compare our predicted emission line intensity ratios to the observed values in our subsets. For each diagnostic diagram, we have adopted the same parameter range, colors, and symbols as in Paper II to enable an easy comparison. In Figs 9-13 below, each set of circles with different colors represents a different density distribution with a power law index (as defined above) in the range $-1.8 \leq \beta \leq -0.6$ in increments of 0.4. The circles within each set represent different $\phi_H$ distributions with power law indices ranging from $-3.5 \leq \alpha \leq -1.5$ in increments of 0.125. This choice of values for $\alpha$ corresponds to the same range in $\phi_H$ that we used in Paper II to describe AGN. The largest circle in each set represents the least negative $\alpha$ value (flattest distribution of incident ionizing flux $\phi_H$). The white circles represent our best fit to each subset and are described below. The observations from the subsets along the central SF locus are also given as black squares on these diagrams, with the largest black square representing the highest ionization case. We have chosen to only show the central locus on these diagrams since the wing loci are very similar to the central locus, as can be seen in Fig. 3.

Each of the 23 different diagnostic diagrams is on the same scale to allow quick comparisons to be made. An ideal fit between the theoretical predictions and the observations would have each of the white circles lying on top of a different black square. As with Paper II, we consider up to a factor of 2 difference as an acceptable fit between our predicted values and the observations. For intensity ratios involving stronger lines, this criterion represents mainly the uncertainty in how well the models describe what must actually be a very complex physical structure which we represent with the still very simplified assumption that the incident flux and density distribution functions can be represented by power laws. The ionizing SED and metallicity presumably also vary from point to point within a galaxy and between different galaxies, as opposed to the standard representative values that we have used here. We have also used the estimated errors discussed in §2 to compute the characteristic observational error bars shown in the upper right corner of each individual panel. The observational uncertainty is significantly smaller than a factor of two for intensity ratios involving only strong lines, but is can be greater than the uncertainties due to the models for ratios involving the weakest emission lines. A brief survey of Figs 9-13 shows that our models largely within the allowed factor of two errors, especially at the higher ionization end of the sequence.

The goal of this paper is find a single, tunable physical parameter responsible for the sequence of pure star forming galaxies. The obvious trend from each of Figs 9-13 is that the observed sequence is well-described by a progressive change in the distribution of the incident ionizing flux $\phi_H$ striking a sea of clouds spread throughout the galaxy. There is no clear evidence for a change in the density distribution $\beta$. The large number of diagnostic diagrams, many of which include weaker emission lines for consistency checks, emphasize the robust nature of this result.

The density distribution parameter $\beta$ is not well constrained for two reasons. First, as $\beta$ decreases the parameter becomes degenerate with each sequence of $\alpha$ lying on top of one another. This can easily be seen by comparing the green and blue sequences of constant $\beta$. Subsequently smaller $\beta$ values generate sequences that lie on top of the blue sequence. Second, a few of the lower ionization emission line ratio subsets correlate with smaller $\beta$ values (e.g. [Ar IV] $\lambda4711$/ [Ar III] $\lambda7135$, He II $\lambda4686$/ H$\beta$, [O III] $\lambda4363$/ [O III] $\lambda5007$) while other emission line ratios (e.g. [N II] $\lambda6584$/ [O II] $\lambda3727$, [N II] $\lambda5755$/ [N II] $\lambda6584$) correlate with larger $\beta$ values.

**Table 5.**
Emission line ratio predictions for dust-free LOC model with only $\alpha$ varied

| | Free Parameters | | | | | Free Parameters | | | | | |
|---|---|---|---|---|---|---|---|---|---|---|---|
| Density Weighting $\beta$ | -1.8 | -1.8 | -1.8 | -1.8 | -1.8 | -1.8 | -1.8 | -1.8 | -1.8 | -1.8 | $\log_{10}$ |
| Ionizing Flux Weighting $\alpha$ | -3.125 | -3.000 | -2.625 | -2.375 | -2.250 | -3.125 | -3.000 | -2.625 | -2.375 | -2.250 | Char. |
| | $\log_{10}$ (predicted line ratio) | | | | | $\log_{10}$ (predicted / observed) line ratio | | | | | Uncert. |
| Line ratio | s01 | s11 | s21 | s31 | s41 | s01 | s11 | s21 | s31 | s41 | |
| [O II] 3727/ [O III] 5007 | 0.85 | 0.77 | 0.42 | 0.08 | -0.16 | -0.08 | -0.08 | -0.20 | -0.16 | -0.15 | 0.20 |
| [O II] 3727/ [O I] 6300 | 1.32 | 1.32 | 1.35 | 1.39 | 1.42 | -0.14 | -0.27 | -0.28 | -0.21 | -0.22 | 0.26 |
| [O II] 3727 / [O II] 7325 | 1.98 | 1.97 | 1.94 | 1.89 | 1.85 | 0.21 | 0.16 | 0.06 | 0.11 | 0.17 | 0.40 |
| [Ne III] 3869 / [O II] 3727 | -1.59 | -1.55 | -1.38 | -1.14 | -0.95 | -0.37 | 0.09 | 0.26 | 0.19 | 0.03 | 0.18 |
| [Ne III] 3869 / He II 4686 | 0.99 | 1.02 | 1.14 | 1.28 | 1.38 | <-0.29 | <0.50 | <-0.19 | -0.01 | -0.07 | 0.38 |
| [S II] 4070 / [S II] 6720 | -1.25 | -1.25 | -1.24 | -1.21 | -1.20 | >-0.25 | >0.26 | 0.07 | 0.19 | 0.23 | 0.40 |
| [O III] 4363 / [O III] 5007 | -2.50 | -2.48 | -2.43 | -2.39 | -2.36 | >-0.93 | >-0.60 | >0.39 | -0.30 | -0.44 | 0.32 |
| He II 4686 / H$\beta$ | -2.29 | -2.29 | -2.25 | -2.21 | -2.18 | >0.01 | >-0.46 | >0.27 | -0.05 | -0.13 | 0.32 |
| He II 4686 / He I 5876 | -1.36 | -1.35 | -1.32 | -1.28 | -1.26 | >-0.32 | >-0.62 | >0.21 | -0.09 | -0.15 | 0.63 |
| [Ar IV] 4711 / [Ar III] 7135 | -2.81 | -2.63 | -1.97 | -1.43 | -1.11 | >-2.89 | >-2.43 | -1.36 | -0.54 | -0.21 | 0.46 |
| [O III] 5007 / H$\beta$ | -0.57 | -0.48 | -0.15 | 0.14 | 0.32 | 0.17 | 0.03 | 0.03 | -0.09 | -0.08 | 0.10 |
| [O III] 5007 / [Ar III] 7135 | 0.62 | 0.70 | 1.00 | 1.28 | 1.46 | -0.65 | -0.36 | -0.17 | -0.16 | -0.08 | 0.26 |
| [N I] 5200 / H$\alpha$ | -2.55 | -2.56 | -2.62 | -2.72 | -2.82 | -0.74 | -0.55 | -0.52 | -0.71 | -0.64 | 0.34 |
| [N I] 5200 / [N II] 6584 | -1.81 | -1.82 | -1.84 | -1.87 | -1.90 | -0.40 | -0.26 | -0.29 | -0.59 | -0.63 | 0.46 |
| [N I] 5200 / [O II] 7325 | -0.39 | -0.40 | -0.48 | -0.57 | -0.65 | -0.60 | -0.33 | -0.28 | -0.33 | -0.22 | 0.63 |
| [N II] 5755 / [N II] 6584 | -2.17 | -2.17 | -2.15 | -2.12 | -2.10 | >-0.12 | >-0.28 | >0.47 | 0.00 | 0.14 | 0.38 |
| He I 5876 / H$\beta$ | -0.94 | -0.94 | -0.93 | -0.92 | -0.92 | 0.33 | 0.16 | 0.06 | 0.03 | 0.02 | 0.40 |
| [O I] 6300 / [O III] 5007 | -0.46 | -0.55 | -0.93 | -1.31 | -1.58 | 0.06 | 0.19 | 0.08 | 0.05 | 0.06 | 0.26 |
| [O I] 6300 / H$\alpha$ | -1.50 | -1.51 | -1.56 | -1.64 | -1.73 | 0.21 | 0.20 | 0.09 | -0.05 | -0.03 | 0.19 |
| [O I] 6300 / [N II] 6584 | -0.76 | -0.77 | -0.78 | -0.80 | -0.81 | 0.55 | 0.49 | 0.32 | 0.06 | -0.02 | 0.24 |
| H$\alpha$ / H$\beta$ | 0.47 | 0.47 | 0.47 | 0.47 | 0.47 | 0.02 | 0.02 | 0.02 | 0.02 | 0.02 | 0.22 |
| [N II] 6584 / [O II] 3727 | -0.55 | -0.56 | -0.57 | -0.59 | -0.61 | -0.41 | -0.23 | -0.05 | 0.15 | 0.24 | 0.26 |
| [N II] 6584 / H$\alpha$ | -0.74 | -0.74 | -0.78 | -0.84 | -0.92 | -0.34 | -0.30 | -0.24 | -0.12 | -0.01 | 0.19 |
| [S II] 6716 / [S II] 6731 | 0.08 | 0.08 | 0.07 | 0.07 | 0.07 | -0.05 | -0.05 | -0.05 | -0.04 | -0.04 | 0.24 |
| [S II] 6720 / H$\alpha$ | -0.47 | -0.48 | -0.52 | -0.61 | -0.70 | 0.18 | 0.09 | 0.02 | -0.01 | 0.05 | 0.12 |
| [O II] 7325 / [O I] 6300 | -0.66 | -0.65 | -0.59 | -0.50 | -0.44 | -0.36 | -0.42 | -0.33 | -0.32 | -0.39 | 0.46 |
| [O II] 7325 / H$\alpha$ | -2.17 | -2.16 | -2.14 | -2.14 | -2.17 | -0.14 | -0.22 | -0.24 | -0.38 | -0.42 | 0.34 |
| Fraction of ratios fitted | --- | --- | --- | --- | --- | 0.67 | 0.78 | 0.67 | 0.81 | 0.85 | |

To determine the best fit between our free parameters and the observations, we started with the highest ionization subset (s41) and adjusted the $\alpha$ and $\beta$ values in our LOC model until the greatest number of emission line ratios were fit to within a factor of 2. We gave more weight to satisfactorily fitting strong emission line ratios, such as those found in VO87, as opposed to ratios from weak emission lines with greater uncertainty. We found that $\alpha = -2.250$ and $\beta = -1.8$ (the blue line on Figs 9-13) provided the best fit to the s41 subset. As mentioned above, while smaller $\beta$ values could also yield a satisfactory fit, we choose $\beta = -1.8$ for consistency with the values provided in Paper II. The rest of the s41-s01 sequence is then fit by holding $\beta$ constant and varying $\alpha$, yielding values of $\alpha = -2.25$, $\alpha = -2.375$, $\alpha = -2.625$, $\alpha = -3.000$, $\alpha = -3.125$ for the s41, s31, s21, s11 and s01 subsets, respectively.

Table 5 presents the theoretical emission line predictions from our best dust-free LOC model for each subset, as well as the comparison between those predictions with the values in each subset. In all but the lowest ionization subset, the fits are usually within the level of uncertainty for each emission line ratio. The last column of the table lists the characteristic observational uncertainty for each line ratio (in $\log_{10}$ format), computed from the estimated uncertainties in the measurements of the individual lines and taking into account additional uncertainty due to the reddening corrections for lines that fall far apart in the spectrum. As with Paper II, this indicates that subsets of galaxies differ by a physically meaningful parameter, the distribution of ionizing flux within those galaxies. The diagnostic diagrams in Figs 9-13 are arranged according to what they physically constrain. We will now elaborate on the comparison between our best model and the observations sorted by these physical characteristics.

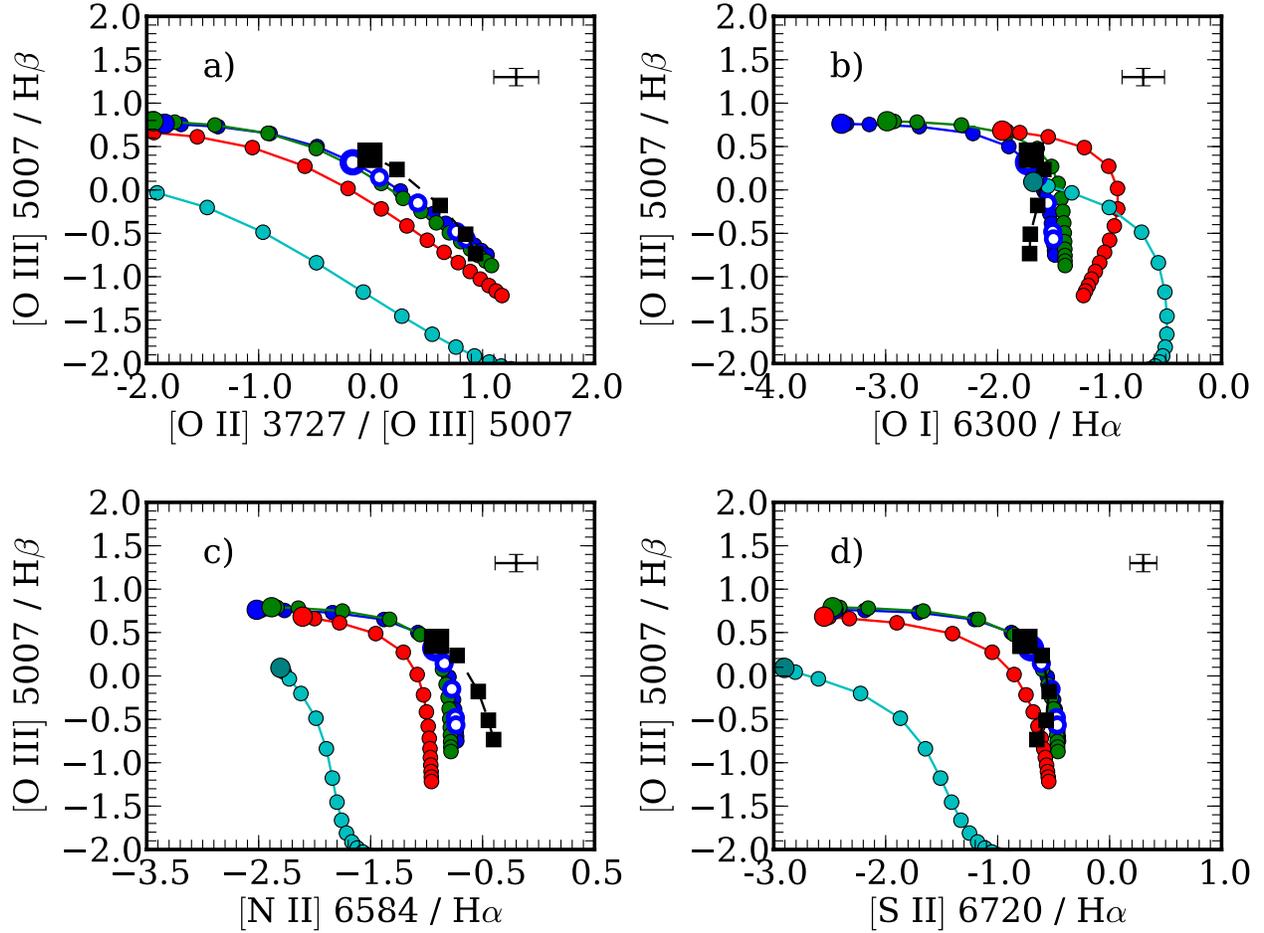

**Figure 9.** Line ratio diagrams constraining the excitation mechanism. Each color represents a different $\beta$ value (density weighting) with cyan: $\beta = -0.6$, red: $\beta = -1.0$, green: $\beta = -1.4$ and blue: $\beta = -1.8$. For each colored line, the individual circles represent different $\alpha$ value (incident ionizing flux $\phi_H$ weighting) in steps of 0.125 with the largest colored circle representing the least negative $\phi_H$ index of $\alpha = -1.5$. The sequence of black squares represents the central set of observations along SF sequence as determined by MFICA with the largest square signifying the highest ionization observation. The open circles in our models indicate the best set of free parameters to fit those observations. The error bars display the uncertainty in each emission line ratio. Our models fit every line ratio along the entire SF sequence except for the lowest ionization case for [N II] $\lambda6584$ / H$\alpha$.

### 3.4.1 Excitation Mechanism

Fig. 9 shows the BPT and VO87 diagnostic diagrams that have been widely successful in constraining the excitation mechanism (either starlight or a power law continuum) in emission line galaxies and therefore constitute the first-order comparison needed to show that MFICA is correctly isolating SF galaxies. Our best models fit the entire sequence for every VO87 diagnostic diagram, except for [N II] $\lambda6584$/ H$\alpha$ in the very lowest ionization subset, to within a factor of 2, and also fit the two highest ionization subsets within the characteristic level uncertainty for each line ratio. This level of agreement between our models and the observations allows us to proceed with extending our analysis to other diagnostic diagrams constraining other properties.

### *3.4.2 Spectral Energy Distribution*

We previously (§3.1) addressed constraining the incident continuum, but we now explore this issue in greater detail. Emission line ratios that use different ions from the same element constrain the SED in the most model-independent way. Due to their high ionization potential, the emission lines He II λ4686 and [Ar IV] λ4711 are particularly useful even when only upper limits can be measured. Fig. 10 shows six different diagnostic diagrams that constrain the SED. In Panel (f) we predict the [S III] λ9069 emission line even though it is outside the range of the observations used here, since it is commonly observed in similar studies. Our models simultaneously fit all of the observational subsets for several of the measured line ratios shown in Fig. 10; specifically these are He II λ4686/ H*β*, [O I] λ6300/ [O III] λ5007, [O II] λ3727/ [O III] λ5007, He II λ4686/ He I λ5876 and [O II] λ7325 / [O I] λ6300.

The [Ar IV] λ4711/ [Ar III λ7135] ratio (panel c) has the greatest discrepancies and only fits the lowest-ionization subsets (due to the observations providing only upper limits), along with the highest ionization subset, but misses for the moderate-ionization subsets. In addition, the [N I] λ5200/ [N II] λ6584 ratio (panel e) fails to fit the higher ionization subsets. All of the strong line ratios, along with ratios incorporating He II λ4686, are fitted very well.

### *3.4.3 Abundances*

Emission lines arising from ions with similar ionization potentials, but from different elements, constrain the gas metallicity due to their insensitivity to the SED. After tuning the gas metallicity according the procedure outlined in §3.2, we used many other diagnostic diagrams to confirm our optimized value of 0.6 Z_solar. All of these diagrams are shown in Fig. 11. The entire He I λ5876 / H*β* and [Ne III] λ3869/ He II λ4686 sequences are fit within their characteristic level of error, while for the [N II] λ6584/ [O II] λ3727 ratio all of the observational subsets are fitted except for the lowest ionization subset.

However, [O II] λ7325 / Hα does not match our models for higher ionization subsets and [O III] λ5007/ [Ar III] λ7135 only matches the high to moderate ionization subsets. Additionally, only the highest ionization subsets are fit for [N II] λ6584 /Hα, which is the subject of discussion in §4.4. The greatest difference between our models and observations comes from [N I] λ5200/ Hα for which none of the subsets are fit.

### *3.4.4 Physical Conditions*

Fig. 12 displays several common diagnostic diagrams constraining the physical conditions of the gas. Panel (a) shows a common temperature indicator [O III] λ4363/ λ5007 against a typical density diagnostic ratio [S II] λ6716/ λ6731. Our models fit the moderate ionization subset and the lowest ionization subsets (to within upper limits) for the [O III] ratio, but under predict this ratio by about a factor of 3 for the highest ionization subset (see panel b). However, we fit the observed [S II] ratio for every subset of to within about 10 per cent.

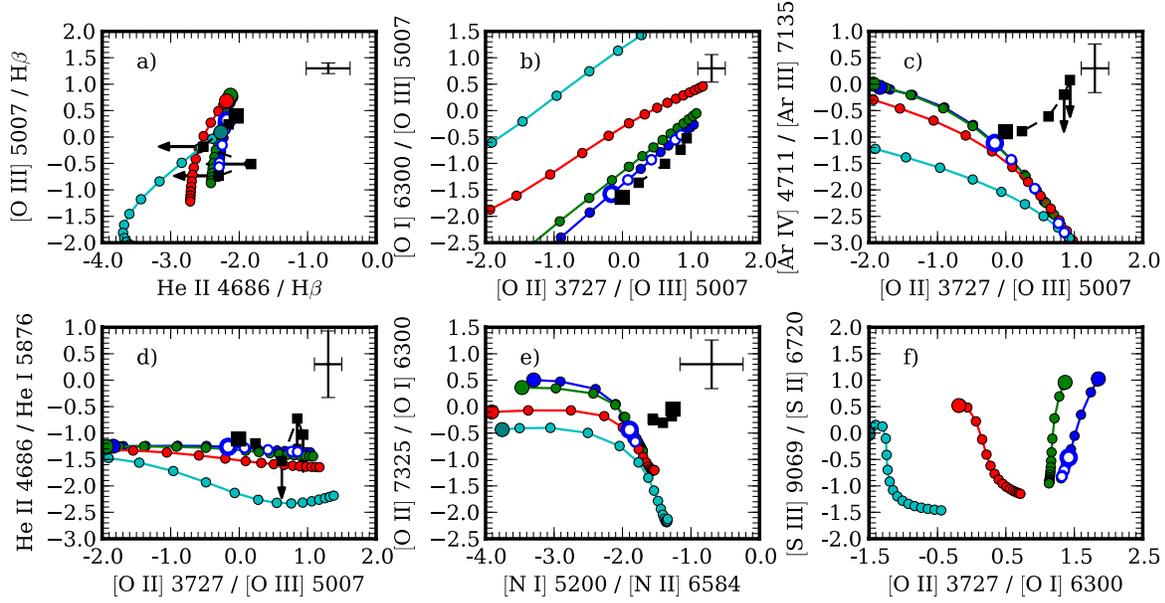

**Figure 10.** Line ratio diagrams constraining the SED, with lines and symbols the same as in Fig. 9. The only discrepancies between our models lie with [Ar IV] λ4711 / [Ar III] λ7135, which only fits the most extreme subset, and [N I] λ5200 / [N II] λ6584 and [O II] λ7325/ [O I] λ6300, which fails to fit all of the observations. All of the other emission line ratios are fit by our models along the entire length of the SF locus.

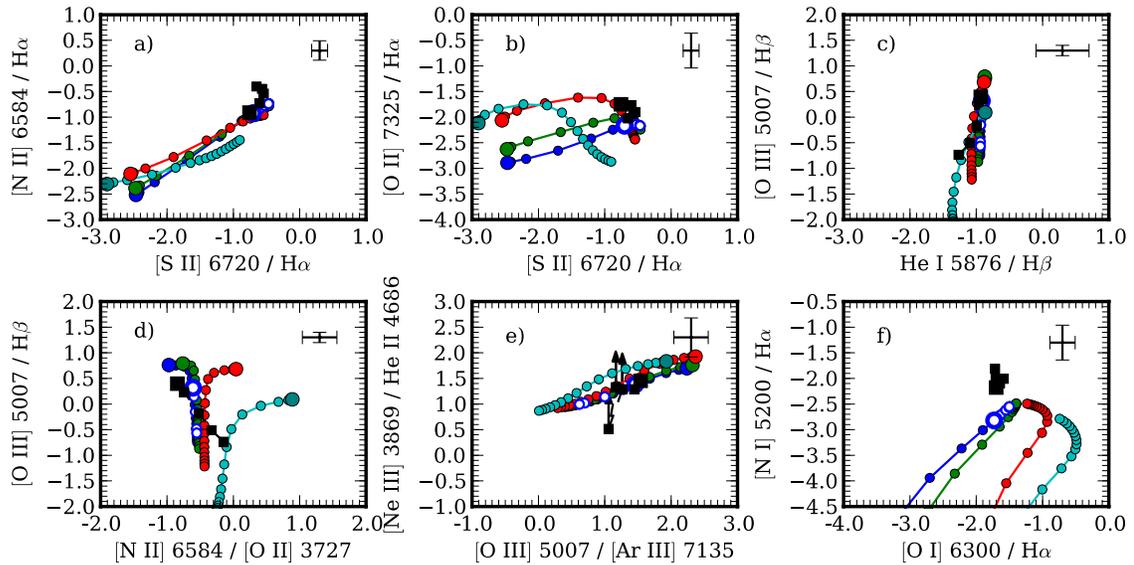

**Figure 11.** Line ratio diagrams constraining abundances, with lines and symbols the same as in Fig. 9. We fit all but the lowest ionization subset for our primary abundance indicator [N II] λ6584/ [O II] λ3727, and also for several other abundance-sensitive emission line ratios.

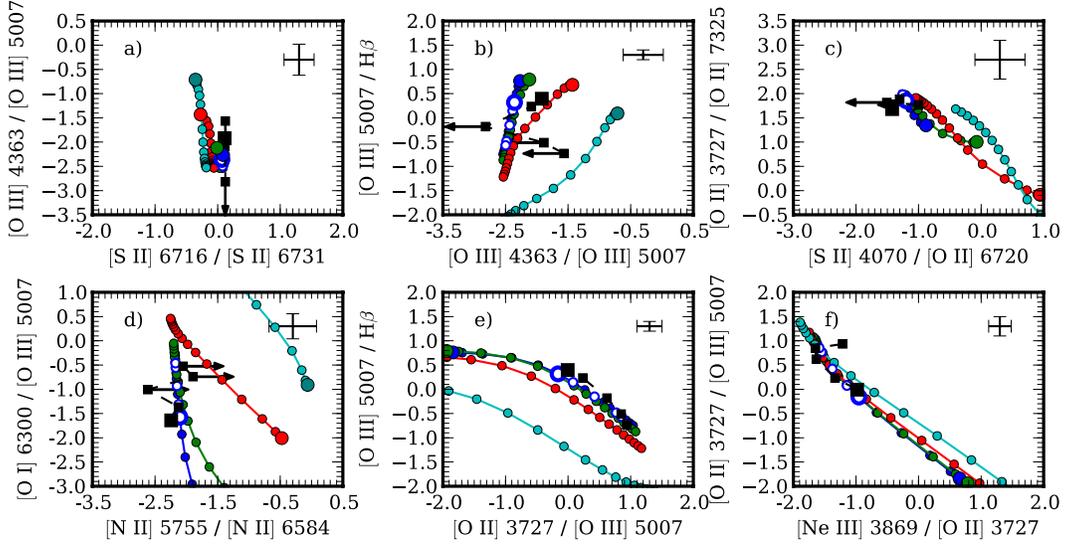

**Figure 12.** Line ratio diagrams constraining the physical conditions, with lines and symbols the same as in Fig. 9. Our models show excellent agreement with the low density diagnostic [S II] λ6716/ λ6731, the high density diagnostics [S II] λ4070/ λ6720 and [O II] λ3727/ [O II] λ7325, along with the ionization parameter diagnostics [O II] λ3727/ [O III] λ5007 and [Ne III] λ3869 / [O II] λ3727. Our models also produce reasonable agreement with the temperature given by the [N II] ratio, but for the highest ionization subsets underpredict the temperature-sensitive [O III] λ4363/ λ5007 ratio.

In addition to the common density diagnostics in Panel (a), our sample size of 50 galaxies per subset allowed us to fit the high density diagnostics shown in Panel (c). Both of the diagnostics, [S II] λ4070/ λ6720 and [O II] λ3727/ λ7325, are fit along the entire SF sequence. While our models do not provide an accurate fit to the observed [O III] ratio, they do fit the temperature sensitive [N II] λ5755/ λ6584 ratio for all of the subsets but one. Finally, Panel (e) repeats the ratio [O II] λ3727/ [O III] λ5007 previously shown in Fig. 9, because this ratio is sensitive to the ionization parameter (Komossa & Schulz 1997) as well as to the ionization mechanism, and our models fit the entire observed sequence. Finally, Panel (f) shows the [Ne III] λ3869 / [O II] λ3727 ratio, which is a good indicator of ionization parameter (Levesque & Richardson 2014), and our models fit the highest ionization and lower ionization subsets.

### 3.4.5 Grain Diagnostics

Groves et al (2004) have proposed a series of line-ratio diagrams as dust indicators, because in their single-cloud models these ratios are sensitive to radiation pressure and/or photoelectric heating from grains. Groves at al. (2004) concluded that dust is needed in order to explain the observed ratios, but here in Fig. 13 we show four of their diagnostic diagrams for our dust-free models. Many of the line ratios used on these diagrams have already been shown here on other diagrams where our models give good fits to the observations. Our dust-free models again give good fits on the dust indicator diagrams in panels (b), (c) and (d) in Fig. 13, which show [O II] λ3727/ [O III] λ5007, [N II] λ6584/ Hα and [O III] λ4363/ [O III] λ5007, all plotted against He II λ4686/ H$\beta$. On the [N I] λ5200/ [O II] λ7325 vs. [O I] λ6300/ [N II] λ6584 diagram in panel (a), the predicted [O I]/ [N II] ratio agrees with the highest ionization subsets but is systematically too high for the lower ionization cases, while the predicted [N I]/ [O II] ratio fits all of the subsets. In this diagram all of the predicted trends are in the same direction as the observed trends, however the predicted [N I]/ [O II] marginally fits within the observational error bars at low ionization despite its high level of uncertainty. Overall, we find that our dust-free models give reasonable agreement with the majority of the observed points on these line ratio diagrams that have previously been proposed as grain diagnostics.

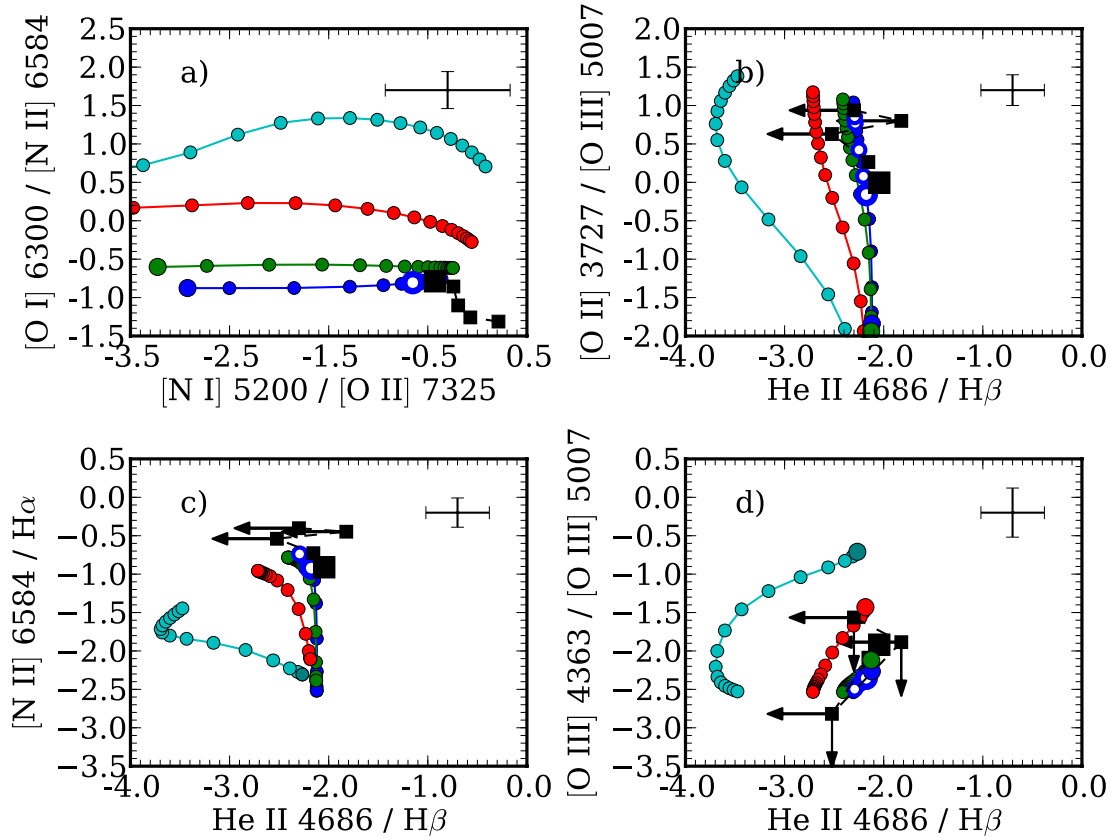

**Figure 13.** Line ratio diagrams proposed by Groves et al (2004) as dust indicators. Our dust-free models are shown, with lines and symbols the same as Fig. 9. Many emission line ratios have been repeated from previous figures, and despite resulting from dust free simulations show nice agreement with the SF locus. The only major discrepancies are that the predicted [N I] λ5200/ [N II] λ6584 ratio disagrees with the highest and lowest ionization subsets while the [O I] λ6300/ [N II] λ6584 ratio only fits the highest ionization subsets.

### 3.5 Dusty Models

Our analysis thus far has been limited to models without dust. Since dust is certainly observed in H II regions, we need to see whether or not including it in our models will change the general conclusions reached from the dust-free models.

At high temperatures, dust will sublimate and thereby enrich the surrounding gas. Rather than attempt to model the complex and poorly-known details of this process, we approximate it with a simple step function. We assume that at $17.9 \leq \log(\phi_H) \leq 20.9$ dust is absent from the gas, which has a solar composition. At $16.9 < \log(\phi_H) \leq 17.9$, we include carbonaceous grains from Orion (Baldwin et al. 1991) and deplete the carbon abundance by 15 per cent. Finally, at $\log(\phi_H) \geq 6.89$ we include both carbonaceous and silicate grain types and assume the abundances of the Orion nebula (Baldwin et al. 1991). While grains of different sizes sublimate at different temperatures, we ignored this effect as we also did in Paper II and just assumed that two step functions in radius were sufficient to understand the results from incorporating dust.

Optimizing the SED for these dusty models resulted in a 5 Myr starburst using the Padova AGB evolutionary track with a continuous SFH, identical to the result for the dust free models. We then used the same procedure as given in §3.2 for optimizing the metallicity. In these dusty models, the gas-phase abundances actually change across the LOC plane due to dust sublimation. Just as with our dust free simulations, the dusty models needed a subsolar metallicity to provide a satisfactory fit. The scaling resulted in a 0.6 Z_solar metallicity for the dust free section of the LOC plane, again the same as was found for the dust-free models.

Table 6 displays the emission line ratio predictions and their comparison to our observations for best fitting dusty model in the same format as Table 5. As a whole, the dusty models are less successful at fitting the s41-s31-s21-s11 subsets but provide a slightly better fit for the lowest ionization subset, s01. This can be seen by comparing the very bottom rows of Tables 5 and 6, which list the fraction of observed emission line ratios that we were able to fit to within their characteristic level of error.

Crucially, our dusty models provide a poor fit to [O I] λ6300/ Hα. They also fit fewer of the subsets for the other strong line ratios (e.g. [S II] λ6720/ Hα, [O III] λ5007/ H*β*, [O III] λ5007/ [Ar III] λ7135). The dusty models fit some of the ratios involving weak emission lines better than the dust-free models do (e.g. [O III] λ4363 / [O III] λ5007), but there are exceptions ([S II] λ4070/ [S II] λ6720 and He II λ4686/ H*β*). Among the ratios for which the dusty models do provide the better fit, [N I] λ5200/ Hα is fit along almost the entire sequence while the dust-free models fail to fit any of the subsets. Our dusty model fits almost the higher ionization [O II] λ7325/ Hα subsets better than our dust-free model. However, our dust-free model provides a better prediction for the ratio of the [N I] λ5200/ [O II] λ7325 emission lines at low ionization.

The emission line ratios presented in §3.4.5 and Fig. 13, which have been proposed by (Groves et al. 2004) to be good indicators of the presence of dust in the particular models that they calculated, do not help distinguish between our two sets of models. Comparison of the values listed in Tables 5 and 6 shows that our dusty models give predictions for the line ratios used on these diagrams that are in most cases within a factor of 2-3 of those for our dust-free models. The two sets of predictions tend to straddle and lie within a factor of two of the observed ratios. This means that the diagrams are not useful discriminants for our models. The actual diagnostic ability of these diagrams seems to depend on the model being applied.

**Table 6.**
Emission line ratio predictions for dusty LOC model with only $\alpha$ varied

| | Free Parameters | | | | | Free Parameters | | | | | $\log_{10}$ Char. Uncert. |
|---|---|---|---|---|---|---|---|---|---|---|---|
| Density Weighting $\beta$ | -1.8 | -1.8 | -1.8 | -1.8 | -1.8 | -1.8 | -1.8 | -1.8 | -1.8 | -1.8 | |
| Ionizing Flux Weighting $\alpha$ | -3.250 | -2.500 | -2.375 | -2.000 | -1.875 | -3.125 | -3.000 | -2.625 | -2.375 | -2.250 | |
| | $\log_{10}$ (predicted line ratio) | | | | | $\log_{10}$ (predicted / observed) line ratio | | | | | |
| Line ratio | s01 | s11 | s21 | s31 | s41 | s01 | s11 | s21 | s31 | s41 | |
| [O II] 3727/ [O III] 5007 | 1.15 | 0.67 | 0.55 | 0.11 | -0.10 | 0.22 | -0.17 | -0.07 | -0.13 | -0.09 | 0.20 |
| [O II] 3727/ [O I] 6300 | 1.30 | 1.33 | 1.34 | 1.39 | 1.42 | -0.16 | -0.26 | -0.29 | -0.21 | -0.22 | 0.26 |
| [O II] 3727 / [O II] 7325 | 1.76 | 1.73 | 1.72 | 1.64 | 1.58 | -0.01 | -0.09 | -0.16 | -0.14 | -0.10 | 0.40 |
| [Ne III] 3869 / [O II] 3727 | -1.74 | -1.57 | -1.51 | -1.22 | -1.05 | -0.53 | 0.08 | 0.13 | 0.12 | -0.07 | 0.18 |
| [Ne III] 3869 / He II 4686 | 1.22 | 1.28 | 1.29 | 1.24 | 1.17 | <-0.06 | <0.76 | <-0.04 | -0.05 | -0.29 | 0.38 |
| [S II] 4070 / [S II] 6720 | -1.08 | -1.07 | -1.06 | -1.02 | -0.99 | >-0.08 | >0.44 | 0.24 | 0.39 | 0.43 | 0.40 |
| [O III] 4363 / [O III] 5007 | -2.21 | -2.15 | -2.12 | -1.94 | -1.78 | >-0.65 | >-0.26 | >0.69 | 0.15 | 0.14 | 0.32 |
| He II 4686 / H$\beta$ | -2.28 | -2.21 | -2.17 | -1.95 | -1.79 | >0.02 | >-0.38 | >0.35 | 0.20 | 0.25 | 0.32 |
| He II 4686 / He I 5876 | -1.36 | -1.30 | -1.27 | -1.07 | -0.92 | >-0.33 | >-0.57 | >0.26 | 0.12 | 0.19 | 0.63 |
| [Ar IV] 4711 / [Ar III] 7135 | -3.12 | -2.24 | -2.05 | -1.45 | -1.21 | >-3.20 | >-2.04 | -1.44 | -0.56 | -0.31 | 0.46 |
| [O III] 5007 / H$\beta$ | -0.47 | -0.04 | 0.06 | 0.39 | 0.52 | 0.27 | 0.48 | 0.25 | 0.16 | 0.12 | 0.10 |
| [O III] 5007 / [Ar III] 7135 | 0.39 | 0.80 | 0.89 | 1.20 | 1.33 | -0.88 | -0.26 | -0.27 | -0.24 | -0.21 | 0.26 |
| [N I] 5200 / H$\alpha$ | -2.03 | -2.12 | -2.16 | -2.38 | -2.51 | -0.22 | -0.11 | -0.07 | -0.37 | -0.34 | 0.34 |
| [N I] 5200 / [N II] 6584 | -1.62 | -1.65 | -1.67 | -1.74 | -1.79 | -0.21 | -0.09 | -0.11 | -0.46 | -0.52 | 0.46 |
| [N I] 5200 / [O II] 7325 | -0.49 | -0.56 | -0.59 | -0.76 | -0.86 | -0.70 | -0.49 | -0.39 | -0.51 | -0.43 | 0.63 |
| [N II] 5755 / [N II] 6584 | -1.89 | -1.87 | -1.87 | -1.83 | -1.80 | >0.17 | >0.02 | >0.75 | 0.29 | 0.44 | 0.38 |
| He I 5876 / H$\beta$ | -0.92 | -0.91 | -0.90 | -0.88 | -0.87 | 0.35 | 0.19 | 0.09 | 0.08 | 0.06 | 0.40 |
| [O I] 6300 / [O III] 5007 | -0.15 | -0.65 | -0.78 | -1.28 | -1.52 | 0.38 | 0.09 | 0.22 | 0.08 | 0.12 | 0.26 |
| [O I] 6300 / H$\alpha$ | -1.08 | -1.16 | -1.19 | -1.37 | -1.49 | 0.63 | 0.55 | 0.45 | 0.21 | 0.21 | 0.19 |
| [O I] 6300 / [N II] 6584 | -0.67 | -0.69 | -0.70 | -0.74 | -0.76 | 0.65 | 0.57 | 0.41 | 0.12 | 0.03 | 0.24 |
| H$\alpha$ / H$\beta$ | 0.47 | 0.47 | 0.47 | 0.48 | 0.49 | 0.01 | 0.02 | 0.02 | 0.02 | 0.03 | 0.22 |
| [N II] 6584 / [O II] 3727 | -0.63 | -0.64 | -0.64 | -0.65 | -0.66 | -0.48 | -0.31 | -0.12 | 0.09 | 0.19 | 0.26 |
| [N II] 6584 / H$\alpha$ | -0.42 | -0.47 | -0.50 | -0.63 | -0.73 | -0.02 | -0.02 | 0.04 | 0.09 | 0.19 | 0.19 |
| [S II] 6716 / [S II] 6731 | 0.07 | 0.06 | 0.06 | 0.05 | 0.05 | -0.05 | -0.07 | -0.06 | -0.06 | -0.06 | 0.24 |
| [S II] 6720 / H$\alpha$ | -0.28 | -0.35 | -0.39 | -0.57 | -0.68 | 0.37 | 0.22 | 0.16 | 0.04 | 0.07 | 0.12 |
| [O II] 7325 / [O I] 6300 | -0.46 | -0.40 | -0.38 | -0.25 | -0.16 | -0.15 | -0.17 | -0.12 | -0.07 | -0.12 | 0.46 |
| [O II] 7325 / H$\alpha$ | -1.54 | -1.56 | -1.57 | -1.62 | -1.65 | 0.48 | 0.37 | 0.33 | 0.15 | 0.10 | 0.34 |
| Fraction of ratios fitted | --- | --- | --- | --- | --- | 0.59 | 0.70 | 0.63 | 0.85 | 0.78 | |

Some of the other line ratios measured here may provide better tests for the presence of dust grains in the context of our models. In particular, the [O I] $\lambda$6300 / H$\alpha$ vs. [N I] $\lambda$5200 / H$\alpha$ diagram (already shown in Fig. 11f, as an abundance indicator) would be very useful. The dust-free models fit the [O I] $\lambda$6300 / H$\alpha$ measurements for almost all of subsets, while the dusty models fit this ratio in none of the subsets. Conversely, the dust-free models fail to fit any of the [N I] $\lambda$5200 / H$\alpha$ subsets yet the dusty models fit the entire locus. This result unfortunately splits the difference between supporting our dust-free or our dusty models, so it is also not definitive.

The main conclusion from our dust-free models was that the systematic changes in the observed line ratios along the SF locus are well-described by a sequence of models in which the distribution in ionizing flux is primary variable. We find that including dust in the models does not affect this result.

## 4. Discussion

### 4.1 Physical Meaning of the SF Locus

For the first time, we have applied an LOC model to understanding SF galaxies ranging from the highest ionization cases down to galaxies very close to the composite region of the BPT diagram. Our large observational sample generated by MFICA provides many weaker emission lines that act as consistency checks for developing a physical understanding of the galaxies. Our goal was to identify the physical parameters responsible for the variation in the emission line properties of galaxies lying upon the SF locus. Figs 9-13 show that this is primarily due to a single physical parameter, the distribution of incident ionizing flux $\phi_H$ striking the individual gas clouds scattered throughout a given galaxy. Indeed, by adopting $\beta = -1.8$ and $\alpha = -2.250, -2.375, -2.625$ our models reproduce almost all of the observed line ratios for the s41-s21 subsets (high to moderate ionization), with only a few exceptions. Simply glancing at Figs 9-13 confirms this result.

Physically, the progressive change in the $\phi_H$ distribution index $\alpha$ means that higher ionization SF galaxies have distributions in $\phi_H$ more heavily weighted towards higher incident ionizing flux. Differences in the $\phi_H$ distribution could in principal be due to differences in how dispersed the star formation activity is within the particular galaxy rather than be the result of differences in the distribution of the ionized gas, or both effects might come into play. Age effects might be important, due to expansion of the individual H II regions over time. Spatially resolved imaging in the UV continuum, H$\beta$, and [O III] of a sample of galaxies covering a wide range along the SF sequence would clarify this issue.

Our biggest discrepancies arise from the lowest ionization subset, s01. The simplicity of our models could account for these differences. We chose to vary the SED, metallicity, and integration weightings, however other criteria could affect the predictions for lower ionization SF galaxies. In particular, adopting different equations of state or more complicated descriptions of the distributions of the cloud density and /or incident ionizing flux could yield different results. Constant pressure simulations have been used in other studies (L10), although in the H II region only a small difference arises from isobaric and isochoric environments since the temperature stays relatively constant until the ionization front.

Additionally, we have assumed that photoionization is the sole excitation mechanism for each subset of galaxies along the SF locus. Indeed, Fig. 9 shows that this is likely to be the case. However, the fits do become worse as one moves towards lower ionization. In the low-ionization galaxies, turbulence, shocks, or cosmic rays could provide additional excitation.

The $E(B-V)$ values in Table 1 show that the lower ionization subsets are more heavily reddened. Our dusty models, although they are as a whole less successful than our dust free models, do fit a few of the lowest ionization subsets slightly better than the dust-free models. This would not be particularly surprising given that the morphology of the low-ionization galaxies tends to indicate a very clumpy, dusty structure and therefore a larger fraction of dusty clouds could affect the emission from the s01 subset. After correcting for extinction, the dereddened [O III] and H$\beta$ luminosities (see Table 1) indicate that the star formation rate (SFR) is about the same in all subsets; i.e. there is no correlation between SFR and the ionization level of the gas. Assuming a Salpeter IMF (Salpeter 1955) and Case B,

$$\text{SFR}(M\_solar\ yr^{-1}) = 2.26 \times 10^{-41}\ L(H\beta) \qquad (3)$$

gives an average SFR ~ 6.0 M_solar yr$^{-1}$ with around a factor of two variation among the subsets (Kennicutt 1998), using our extinction corrected luminosities and assuming $I(H\alpha)/I(H\beta) = 2.86$. Although the derived SFR is sensitive to IMF, we can safely assume that the majority of the galaxies within each subset are undergoing relatively normal rates of star formation. Note that reddening is a major source of uncertainty in estimating SFRs from recombination lines, and the constant SFR feature in our subsets only arises after correcting for extinction. The absence of a correlation with the SFR supports the idea that the distribution of flux surrounding the sites of SF is the primary variable that changes along the SF sequence, rather than it being the secondary result of differences in the nature of the starburst itself.

### 4.2 Limitations in population synthesis models

Our models fit the observed [S II] / Hα ratio, indicating that Starburst99 simulations provide a reasonable prediction for the FUV flux. Yet, problems still exist in the SED at even higher photon energies as can be seen from the increase in [Ar IV] / [Ar III] as one moves from the high ionization subsets to the moderate ionization subsets (Fig. 10c). The noticeable mismatch at moderate ionization implies that a substantially harder EUV continuum is needed, but unfortunately this is not available using the Starburst99 tracks that we considered for the LOC models.

New evolutionary tracks have recently become available that include stellar rotation (Leitherer et al. 2014). The main effect of rotation relates to stellar lifetimes causing main sequence stars to burn hotter and longer with higher luminosities (Levesque et al. 2012). Rotation also promotes mass loss giving rise to a greater fraction of WR stars. The overall net result is a significantly harder spectral energy distribution.

For this reason, we ran a single LOC grid using the Geneva evolutionary track that featured rotation with $v = 0.4$ and metallicity $Z = 0.008$ (Georgy et al. 2013) at an age of 7 Myr. At this age the starburst has reached steady state for a continuous SFH, and the radiation field is significantly harder than with the Geneva high mass loss tracks. We used a subsolar gas metallicity of $Z = 0.6$ Z_solar consistent with our best model. Fig. 14 shows the same diagnostic diagrams used for constraining the SED as Fig. 10, but with the incident continuum described above. The figure indicates that the hard photon problem persists in that the predicted [Ar IV] / [Ar III] does not increase as one moves towards lower ionization subsets. Crucially, this model also fails to correctly predict the He II λ4686 emission line for the higher-ionization cases, which was successfully fitted by the Padova models (Fig. 10). This shows that the non-rotating Padova AGB evolutionary tracks provide a better fitting ionizing spectrum than the rotating Geneva evolutionary tracks for our test case. One possible solution to creating more high energy photons could therefore consist of incorporating rotation into the Padova group AGB evolutionary tracks. Another possible solution could involve incorporating binary stellar evolution into stellar population synthesis models (Stanway et al. 2014).

### 4.3 Comparisons to Previous Work

We next compare the quality of the fits obtained with our LOC models to those obtained with widely used models which describe the SF sequence as being mainly an "abundance sequence" due to a systematic variation in the metallicity (cf. Kewley et al. 2013). We performed our calculations with the spectral simulation code Cloudy described by Ferland et al. (2013), while calculations by L10, Kewley (e.g. Kewley 2001, 2013) used the code Mappings. The two codes are attempting to perform the same calculation, that is, follow the conversion of the stellar SED into an emission line spectrum. The only direct comparison between using these two codes to describe an identical physical situation was presented in Ferland et al. (1995). Due to the similarity of the results from this comparison, we believe that the differences in our predictions are more due to the different assumed geometry and the integrations over ensembles of clouds, than to the codes themselves.

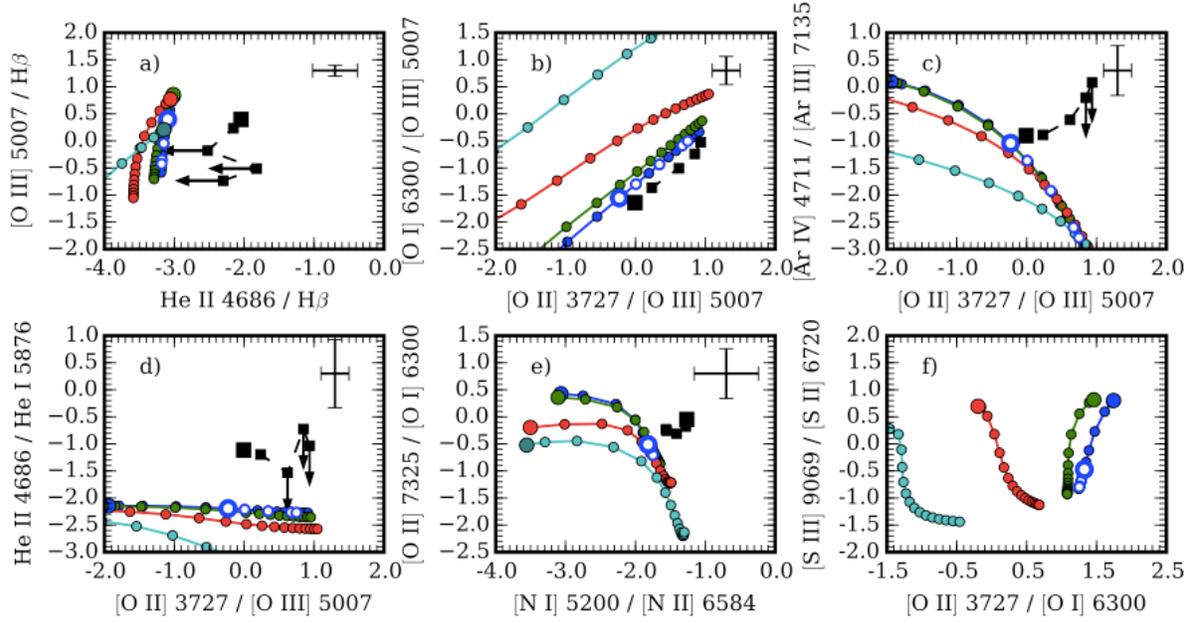

**Figure 14.** Line ratio diagrams constraining the SED as in Fig. 10, with lines and symbols the same as in Fig. 9, but with an incident radiation from the Geneva evolutionary track that features stellar rotation. Panels (a), (c), and (d) show that the hard photon problem still persists since higher ionization emission lines are not satisfactorily fit.

Dopita & Evans (1986) originally published grids of models, which suggested that the sequence of SF galaxies on the BPT diagram could be described by correlated variations in the ionization parameter and metallicity. This work was refined in later papers by Kewley et al. (2001) and L10, which concentrated on evaluating improved stellar atmosphere models to produce SEDs that can better reproduce the observed emission line ratios. Their plasma simulations used constant pressure plane parallel gas clouds with a fixed electron density. Any given galaxy was assumed to be characterized by two parameters. The first is a single ionization parameter, defined as $q = cU$, where $c$ is the speed of light. The second parameter is the metallicity, which L10 designate in most of their diagrams by $z = 0.02\,Z$.

We use the L10 models for our comparison because they employ the most recent SEDs. Although the L10 paper only shows line ratio diagrams for a quite limited set of emission lines, the full computed set of line strengths for their grid of models is publically available on the internet[2]. This data set lets us form all of the same line ratios that we used for our comparisons of the LOC model to the observations. We used L10's continuous star formation, 5.0 Myr models with electron density $n_e = 100$ cm$^{-3}$, which is the preferred case shown in their Figs. 5, 7 and 9. For each of the points s01, s11, s21, s31 and s41 along the SF sequence, we used the [O III] λ5007/ [O II] λ3727 vs [N II] λ6584/[O II] λ3727 line ratio diagram (Fig. 7 in L10) to determine $q$ and $z$ from the observed line ratios. We then interpolated in the L10 model grid to find all of the other line ratios for these $q$ and $z$ points, and calculated the predicted/observed values in the same way that we did for our LOC models. We did this three times, first allowing two free parameters (the ionization parameter $q$ and the metallicity $z$) to vary. Then for our second comparison, we fixed the ionization parameter at $q = 3 \times 10^7$ and varied only $z$. This value of $q$ was chosen so that the average residual in log([O III] λ5007/ [O II] λ3727) was 0. Then for our last comparison, we fixed the metallicity at $z = 0.02$ and varied only $q$. This value of $z$ was chosen so that the average residual in log([N II] λ6584/ [O II] λ3727) was 0.

**Table 7**
Comparison model parameters and results

|  | s01 | s11 | s21 | s31 | s41 |
|---|---|---|---|---|---|
| L10 models, varying $q$ and $z$: | | | | | |
| $q \times 10^{-7}$ | 2.0 | 2.0 | 2.7 | 4.0 | 6.6 |
| $z$ | 0.031 | 0.024 | 0.019 | 0.014 | 0.012 |
| L10 models, fixed $q$, varying $z$: | | | | | |
| $q \times 10^{-7}$ | 3.0 | 3.0 | 3.0 | 3.0 | 3.0 |
| $z$ | 0.031 | 0.024 | 0.019 | 0.014 | 0.012 |
| L10 models, fixed $z$, varying $q$: | | | | | |
| $q \times 10^{-7}$ | 1.6 | 1.9 | 2.7 | 5.4 | 9.3 |
| $z$ | 0.02 | 0.02 | 0.02 | 0.02 | 0.02 |

**Table 8**
Fraction of ratios fitted for the L10 model and our LOC models

|  | s01 | s11 | s21 | s31 | s41 |
|---|---|---|---|---|---|
| **LOC Dust-Free Models:** | | | | | |
| Varying $\alpha$ and $Z$ | 0.56 | 0.74 | 0.67 | 0.85 | 0.93 |
| Fixed $\alpha$, varying $Z$ | 0.63 | 0.67 | 0.63 | 0.74 | 0.63 |
| Fixed $Z$, varying $\alpha$ | 0.67 | 0.78 | 0.67 | 0.81 | 0.85 |
| **LOC Dusty Models:** | | | | | |
| Varying $\alpha$ and $Z$ | 0.63 | 0.74 | 0.59 | 0.89 | 0.81 |
| Fixed $\alpha$, varying $Z$ | 0.67 | 0.74 | 0.59 | 0.74 | 0.63 |
| Fixed $Z$, varying $\alpha$ | 0.59 | 0.70 | 0.63 | 0.85 | 0.78 |
| **L10 Models:** | | | | | |
| Varying $q$ and $z$ | 0.63 | 0.74 | 0.63 | 0.44 | 0.44 |
| Fixed $q$, varying $z$ | 0.63 | 0.67 | 0.63 | 0.52 | 0.52 |
| Fixed $z$, varying $q$ | 0.70 | 0.78 | 0.67 | 0.44 | 0.37 |

Table 7 lists the adopted model parameters for the L10 described above and our LOC models. The "fraction of ratios fitted" listed in Table 8 show that the L10 models in which both $q$ and $z$ are allowed to vary fit the observations approximately the same as the L10 models in which only $z$ is allowed to vary. However, our LOC dust-free models fit the observed SF sequence at least as well, and arguably better than, either sequence of L10 models.

---

[2] http://www.emlevesque.com/model-grids/

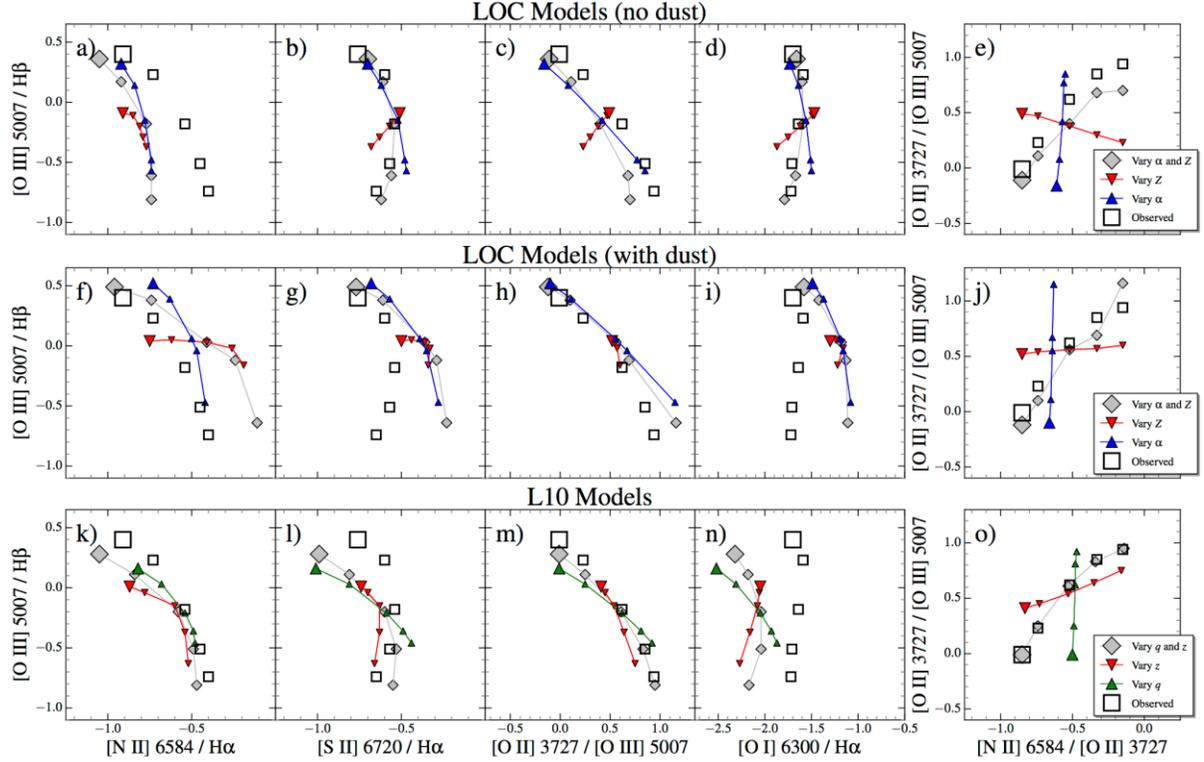

**Figure 15.** The four VO87 diagnostic diagrams and the [N II] / [O II] ratio for our LOC models (top panels) and the L10 models (bottom panels). The LOC and L10 models are each shown for three different cases: varying two parameters, and then varying each parameter individually. The values for metallicity were interpolated to fit the [N II] / [O II] ratio, a common metallicity indicator. In the LOC model, the orthogonality to $\alpha$ and smaller range of predicted line ratios shows that our free parameters are not as degenerate as those in the L10 model and that metallicity is a secondary, not primary, physical parameter.

For a more in depth comparison, Fig. 15 shows the VO87 diagnostic diagrams and the [N II] $\lambda 6584$ / [O II] $\lambda 3727$ ratio for the L10 models and our LOC models. The top panel displays three sets of LOC models. First, we allowed both $\alpha$ and $Z$ to vary. Next, we fixed $\alpha = -2.625$, which is the best fitting value for the moderation ionization subsets in Fig. 9-13, and interpolated $Z$ to fit the [N II] $\lambda 6584$ / [O II] $\lambda 3727$ ratio. Finally, we also show the same models given in Fig. 9-13 where $Z = 0.6$ and $\alpha$ is allowed to vary.

Our models predict only half of the observed factor three change in the frequently used (and reddening-free) [N II] $\lambda 6584$ / H$\alpha$ ratio, while all three sets of L10 models span the observed level of variation. The L10 models have a slight discrepancy in fitting the [S II] $\lambda 6720$ / H$\alpha$ ratio, in particular for higher ionization galaxies, in two sets of their models. We have not encountered this problem here and find that our models satisfactorily fit this diagnostic ratio when varying $\alpha$ and $Z$, along with varying only $\alpha$. Previously, deficits in models failing to fit the [S II] $\lambda 6720$ /

Hα ratio were attributed to a lack in FUV photons, however our LOC models show that discrepancy could also be a function of the underlying model that is assumed instead of problems with the incident radiation field.

A bigger problem with the L10 models is the line ratios involving [O I] λ6300, where again our LOC models do much better. The difference in the equation of state used in the L10 and our LOC models (constant pressure vs. our constant density) might be responsible for a small amount of the difference in [O I]. Another problem with the L10 models comes from fitting [O III] λ5007 with their preferred abundance sequence model. When only varying Z, the L10 models fail to fit the highest ionization observations. Our models show a similar discrepancy when only varying Z, however we discuss below that metallicity is a secondary parameter in our models while it is a primary parameter in the L10 models. Thus, the primary parameter in the L10 models cannot account for the higher ionization [O III] λ5007 observations that are indicative of ionization level.

Finally, while our models do fit the metallicity sensitive [N II] λ6584 / [O II] λ3727 ratio to within the uncertainty in 4 out of 5 cases while holding Z constant, they do not predict the observed variation in this ratio along the SF locus (Fig. 11d). They predict only a 0.06 dex change in this ratio as the $\phi_H$ distribution parameter changes, while the observations show a steady 0.7 dex increase in the ratio when moving to progressively lower ionization star forming galaxies. If α is held constant in our LOC model, but Z is interpolated to fit the [N II] λ6584 / [O II] λ3727 ratio (red triangles), our models only require the metallicity to span Z = 0.4 (low ionization) to Z = 0.94 (high ionization). This is a factor of ~2.5 variation in metallicity while the L10 models similar level of variation.

In addition, it is worth noting two other effects of metallicity in our LOC models that differ from the L10 models. First, varying only metallicity produces a much smaller range in line ratio values for our models than for the L10 models. In Fig. 15, the red triangles for the LOC models cover a smaller part of each diagram compared to the red triangles for the L10 models. In particular, models with only variable metallicity are only able to cover a small range of ionization, -0.37 < log [O III λ5007] / Hβ < -0.09. This emphasizes even further that metallicity is a secondary parameter. Second, varying only the metallicity in our models produces a locus of points that is orthogonal to the locus of observed points on the diagrams plotting [O III] λ5007 /Hβ vs [S II] λ6720 / Hα, [O II] λ3727 / [O III] λ5007, and [O I] λ6300 / Hα (Figs 14b,c,d). For the L10 models, the cases where metallicity, ionization parameter, or both are varied all show similar trends. Thus, our parameters are more independent of each other, while the L10 parameters tend to become degenerate. We interpret this to mean that in our models metallicity is a secondary parameter describing the SF sequence but that much of the metallicity effect seen in the L10 models has been soaked up by our more realistic description of the distributions of the incident ionizing flux $\phi_H$ and the gas density $n_H$.

Aside from differences in common strong line ratios, there are other notable differences in comparing our models to L10 that contribute to the aggregated information in Table 8. One major difference between our LOC models and the L10 models is for the high-ionization He II 4686 line, which in fact is detectable in the observed spectra of the higher ionization subsamples (s31 and s41). The L10 models underpredict the He II line by a factor of 60 while our LOC models properly predict the observed lines. The LOC models also do somewhat better than the L10 results for [Ar IV] λ4711, underpredicting the strength of this line by factors of 23 and 3 for the s21 and s31 subsamples, respectively, as opposed to factors of 160 and 29 with the L10

| | Table 9 | | | | | | | | | |
|---|---|---|---|---|---|---|---|---|---|---|
| | Predicted/observed line ratios for Moy et al (2001) models | | | | | | | | | |
| | Varying $f$ and $Z$ | | | | | Constant $f$, varying $Z$ | | | | |
| Parameter: | s01 | s11 | s21 | s31 | s41 | s01 | s11 | s21 | s31 | s41 |
| $\log_{10} f$ | -4.00 | -3.72 | -3.34 | -2.90 | -2.50 | -3.25 | -3.25 | -3.25 | -3.25 | -3.25 |
| $Z$ | 0.87 | 0.81 | 0.69 | 0.50 | 0.42 | 0.87 | 0.81 | 0.69 | 0.50 | 0.42 |
| $\log_{10}$(predicted/observed) line ratio: | | | | | | | | | | |
| [O III] 4363 / [O III] 5007 | | | | -0.47 | -0.50 | | | | -0.47 | -0.50 |
| [O III] 5007 / Hβ | 0.01 | -0.01 | 0.00 | -0.02 | -0.01 | 0.45 | 0.28 | 0.06 | -0.21 | -0.34 |
| [O I] 6300 / Hα | -0.01 | 0.00 | 0.00 | 0.03 | 0.01 | -0.13 | -0.08 | -0.02 | 0.12 | 0.26 |
| [N II] 6584 / Hα | 0.19 | 0.23 | 0.31 | 0.40 | 0.51 | 0.17 | 0.22 | 0.32 | 0.45 | 0.56 |
| [S II] 6720 / Hα | 0.11 | 0.02 | -0.01 | 0.02 | 0.01 | 0.02 | -0.04 | -0.03 | 0.10 | 0.23 |
| [O II] 3727/Hβ | -0.11 | -0.17 | -0.11 | 0.02 | 0.05 | -0.01 | -0.11 | -0.10 | 0.03 | 0.14 |
| Fraction of ratios fitted, M01 | 1.00 | 0.60 | 0.60 | 0.67 | 0.67 | 0.60 | 0.40 | 0.60 | 0.50 | 0.00 |
| Fraction fitted, LOC dust-free | 0.00 | 0.40 | 0.60 | 1.00 | 0.83 | 0.00 | 0.40 | 0.60 | 1.00 | 0.83 |

models, and successfully predicting this line for the s41 subsample while the L10 model underpredicts it by a factor of 10. However, this may be due to the differences between the SEDs used in the two sets of models rather than to the difference between varying the gas distribution as opposed to varying the metallicity.

Another study similar to that of L10 and Kewley et al (2001) is by Moy, Rocca-Volmerange & Fioc (2001; hereafter M01). They varied the initial IMF, age and metallicity $Z$ of an instantaneous starburst and $Z$ and the filling factor $f$ (serving as a surrogate for the ionization parameter $U$) of the ionized gas. M01 argued that the apparent link between metallicity and ionization parameter is better described as an increase in the dispersion of $U$ at low metallicity $Z$, which they suggested was due to the dominance of line emission from low-excitation H II regions in more evolved (higher metallicity) starbursts. We do find a change in the dispersion of the line ratios along our SF locus, showing up as an increase in the separation of the wing sequences indicated by the blue triangles and diamonds in Fig. 3 and the size of the ellipses in Fig. 4, but the dispersion tightens up at each end of the SF ionization sequence.

Here we simply compare the points along our central SF locus to the M01 grids of predicted line ratios as a function of filling factor $f$ vs. metallicity $Z$. We used our observed line ratios for each point along the SF sequence to find $f$ and $Z$ using the M01 [O I] λ6300/ Hα vs [O III] λ5007/ Hβ diagram from their figure 1, and then we used these $f$, $Z$ points to find predicted values for the remaining line ratios shown on the diagrams in their paper. We then repeated the exercise for a fixed filling factor $\log f$ = -3.25, chosen to minimize the line ratio residuals on the [O I] λ6300/Hα vs. [O III] λ5007/Hβ diagram, and using the same metallicities as determined previously. The resulting log(predicted/observed) values are listed in Table 9, together with some the model parameters that specify each point along the SF sequence. The final two rows in the table show the fraction of line ratios fitted within their level of uncertainty, for the M01 models and then for our LOC dust-free models using the same line ratios. Note, the M01 grids do not cover the parameter space needed to determine the predicted [O III] λ4363 / [O III] λ5007 ratio for the s01, s11 and s21 points, so we have left those entries blank.

The results in Table 9 are similar to those for the L10 comparison, in the sense that for the rather small number of line ratios that are available in M01, our LOC models agree with the observations at least as well as do the M01 models. Neither set of models fits the observed [O III] λ4363 / [O III] λ5007 ratio. Our LOC models do not correctly fit the observed [N II] λ6584 /

Hα ratio for the moderate to low ionization cases, while the M01 models only fit this ratio for the lowest ionization case. We note that it is not possible to precisely recover the M01 predictions of the [N II] λ6584 / Hα ratio due to the very dense overlying set of observed data points on the diagram in their paper, but the model grid lines can be followed well enough to see that the observed ratios for our SF sequence are systematically considerably smaller than the ratios predicted using the model parameters that were determined from the much cleaner [O I] λ6300/ Hα vs [O III] λ5007/ Hβ diagram.

The above comparisons show that the interpretation of the SF sequence as mainly an abundance sequence is not unique. The SF sequence can be described at least as well as principally a sequence in the relative concentration towards high values in the distribution of the ionizing flux $\phi_H$ incident on the individual gas clouds scattered throughout a complicated SF galaxy, as in our LOC models, and in fact this latter interpretation provides a somewhat better fit to the observed line ratios. However, even with our interpretation additional considerations are still needed to best fit the observed [N II] line strengths.

## 4.4 Addressing Discrepancies Between Predictions and Observations

The top row of Fig. 15 shows that the dust-free LOC models provide good fits to all of the strong-line ratios except that they underpredict the [N II] λ6584 / Hα ratio for all except the highest ionization galaxies. A number of physical properties could account for this mismatch. Varying the metallicity does not solve this problem, although it does allow our models to fit the observed [N II] λ6584 / [O II] λ3727 ratio. This can be compared to the L10 models shown in the bottom row of the Figure, which do fit the [N II] λ6584 / Hα ratio, but systematically underpredict [O I] λ6300 / Hα across the full ionization range and also underpredict [S II] λ6720 / Hα for the high ionization cases.

The dusty LOC models in which only α is varied, shown as triangles in the middle row of Fig. 15, correctly follow the observed [N II] λ6584 / Hα trend (Fig. 15f). However, at the low-ionization end of the SF sequence they overpredict [S II] λ6720 / Hα and [O II] λ6720 / Hα, and they do not predict the observed variation in [N II] λ6584 / [O II] λ3727. Allowing both α and the metallicity to vary in these dusty LOC models (the diamond symbols in the Figure) provides a good fit to [N II] λ6584 / [O II] λ3727 (Fig. 15j), but at the cost of now overpredicting the [N II] λ6584 / Hα ratio for low-ionization galaxies.

Table 8 shows the fraction of ratios fitted for all of these different models, but now including all of the ratios involving weaker lines. This again shows that at the high-ionization end of the SF sequence, the dust-free LOC models fit the observations better than the L10 models, but for the lowest-ionization case (s01) the L10 models fit marginally better. They also show that on balance, the dusty LOC models fit the low-ionization cases about as well as do the L10 models, but fit better than the L10 models at the high-ionization end of the sequence.

The fact that the dusty models fit the low ionization part of the SF locus slightly better than the dust free models leads one to believe that lower ionization galaxies have greater dust abundances. Indeed, our E(B-V) values given in Table 1 confirm that ionization level is inversely correlated with reddening. As described in §3.5, the dust-indicator diagrams proposed by Groves et al. (2004) do not separate out our dust-free from our dusty models, and the [O I] λ6300 / Hα vs. [N I] λ5200 / Hα diagram gives results that fall between our two sets of models.

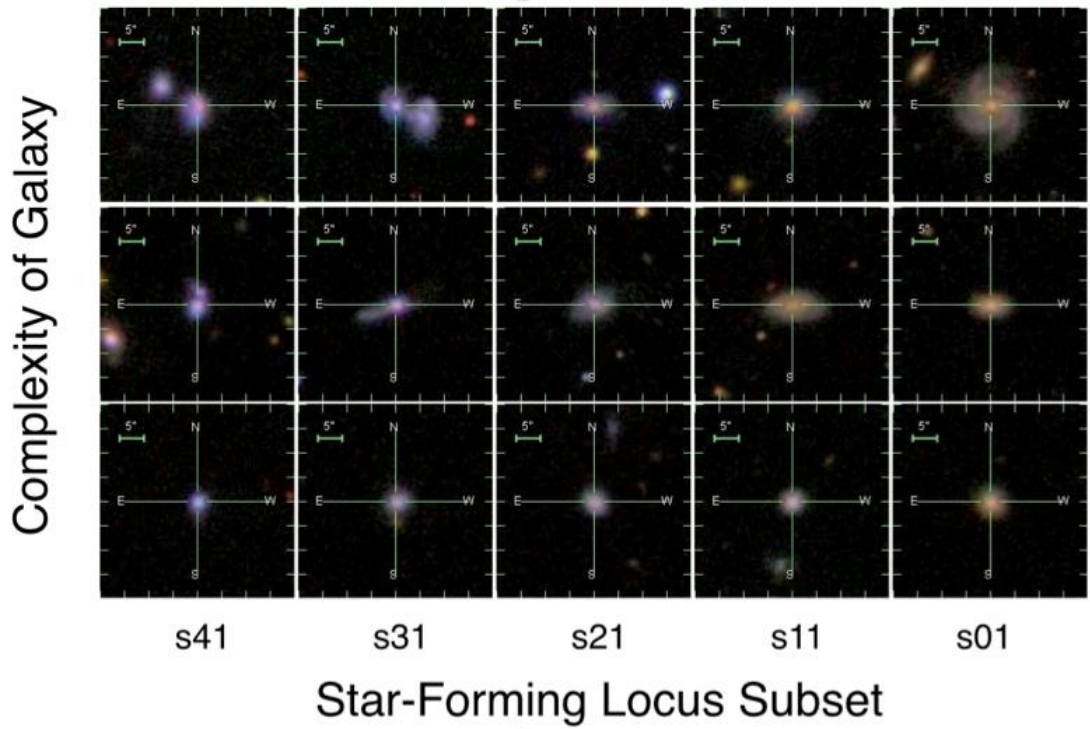

**Figure 16.** Typical morphologies for our SDSS galaxy sample. Galaxies in the higher ionization subsets (s41, s31) are often have irregular morphologies or are blue compact galaxies, while galaxies in the lower ionization subsets (s11, s01) are mostly spiral galaxies.

This argues for a variety of cloud compositions within a single galaxy, which is natural to envision. Both the L10 model and our LOC models assume a single abundance set within each galaxy, but in reality local H II regions do not have identical compositions. Incorporating clouds with both low dust to gas ratios and high dust to gas ratios within a single simulation grid could possibly reconcile the differences between these two extreme cases.

This line of reasoning leads to a physical picture where high ionization SF galaxies contain H II regions that have far lower dust to gas ratios than low ionization SF galaxies, which contain H II regions sprinkled with a variety of dust to gas ratios. We briefly examined the morphological structure of our galaxy sample in Fig. 16, which displays three characteristic galaxies of varying complexity for each of the central subsets in our sample. The more highly ionized galaxies (s41, s31) are predominantly mergers, or the product of mergers, while the less ionized galaxies (s11, s01) are spirals.

At this distance, we cannot resolve individual H II regions, even those as large as 30 Doradus or NGC 604. However, observations of nearby galaxies suggest that when compared to spirals, irregulars should have are larger fraction of their volume filled with H II regions (Youngblood & Hunter 1999, Kennicutt 1984), which is consistent with our finding that highly-ionized star forming (s41) galaxies have more high energy photons ($\alpha = -2.250$) compared to weakly-ionized star forming (s01) galaxies ($\alpha = -3.125$). Indeed, observations of local galaxies confirm that irregulars or blue compact galaxies (s41, s31) have lower dust to gas ratios than spirals (s11, s01) (Rémy-Ruyer et al. 2014) possibly due to the shocks resulting from gas-rich galaxy mergers, which can destroy dust through grain sputtering (Draine & Salpeter 1979).

The two distinct populations of galaxies, one related to mergers and one related to local spirals,

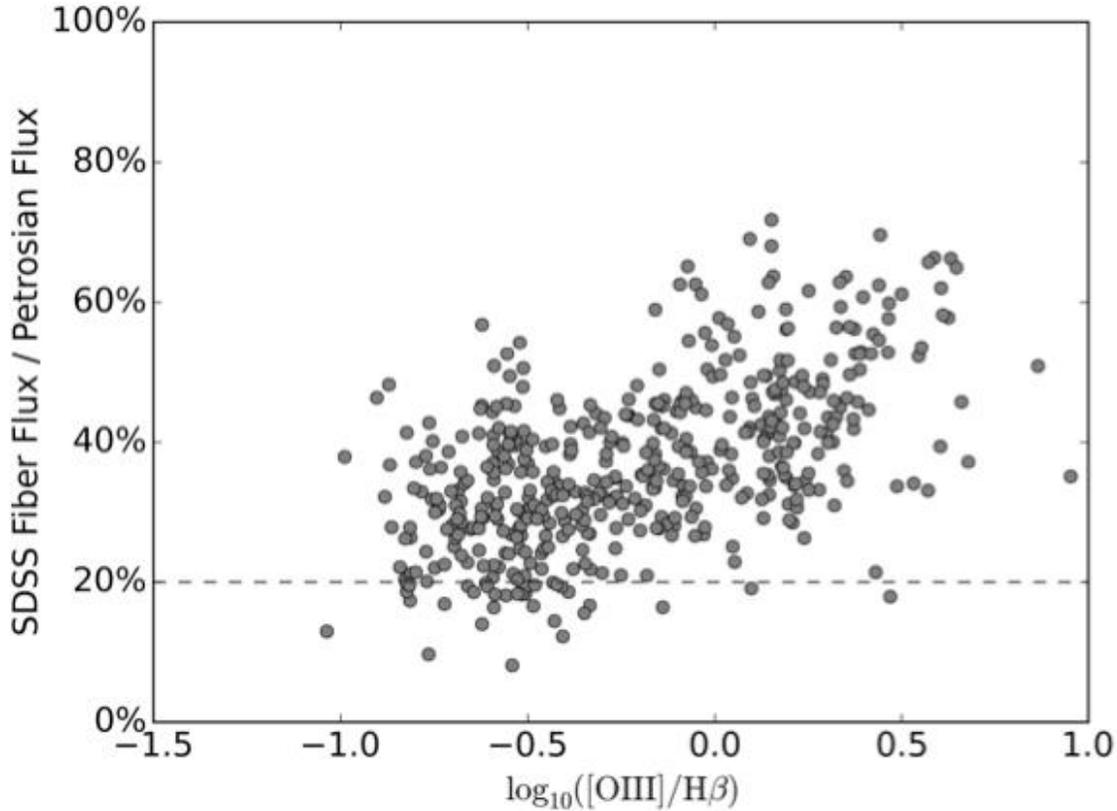

**Figure 17.** The fraction of light captured by the fiber in the r-band relative to the ionization level for the galaxies used to form our composite spectra. Despite the correlation between ionization level and the galaxy size, according to the Kewley et al. (2005) criterion the aperture losses due to the 3" SDSS fiber should not strongly affect capturing the global properties of each galaxy even at the lowest-ionization.

might require different physical descriptions of the gas in an LOC model. In our models, and as in other papers that apply an LOC model, we assumed a spatial distribution function where the distributions of density and ionizing flux are described by power laws for simplicity. Power law distributions have been successful in modeling AGN where a central ionizing source illuminates surrounding clouds that produce emission lines. This physical picture is similar to the blue compact galaxies that occupy the s41 subset where a central starburst excites the surrounding medium, and for which our models fit the emission line spectra very well. In a spiral galaxy, such as those found in our s01 subsets, the sources of ionization are scattered throughout the spiral arms. It is possible that these lower ionization galaxies could feature a different form of spatial distribution function or that a different form of spatial distribution function could describe the entire SF locus. For example, a spatial distribution function in the form of fractals has been used to fit quasar spectra (Bottorff & Ferland 2001). Investigating these effects is beyond the scope of this paper, however we intend to pursue this avenue further in a future paper.

The key factor for the analysis in this paper is the large spatial scale that the SDSS fibres include at z ~ 0.11. It does not matter if the SDSS spectra capture the entire light from the galaxies as long as they include spatial scales that produce representative contributions to the total emission line spectra (i.e. as mentioned at end §3.3). As was noted in §2, Kewley et al. (2005) and Brinchmann et al. (2004) found that an entrance aperture that captures 20 per cent of the total light is sufficient for this purpose. Fig. 17 compares the fraction of light captured by the fiber in the r-band to the ionization level for the galaxies used to form our composite spectra. While there is a correlation between ionization level and the galaxy size, according to the Kewley et al. (2005) criterion the aperture losses due to the 3" SDSS fiber should not strongly affect the results for even the lowest-ionization galaxies. However, if for some reason there are much stronger radial gradients in ionization level within the individual galaxies in our sample than were present in the sample studied by Kewley et al. (2005) and Brinchmann et al. (2004), the fiber size could still introduce a correlation between ionization level and galaxy size. But even so, this would affect any paper analyzing these SDSS spectra, so our basic conclusion would remain valid: the trends in the *measured* emission-line spectra along the SF sequence may well be due more to a difference in the ionizing flux distribution than to metallicity variations.

## 5. Conclusions

As we did in Paper II for AGN, we have used LOC models to understand the physical parameter responsible for the ionization sequence of SF galaxies. Our analysis indicates that by assuming a single shape for the SED, and a single gas metallicity, a varying distribution of the ionizing flux $\phi_H$ incident on the individual clouds can fit several emission line ratios that constrain the excitation mechanism, abundances, SED, and physical conditions, over a wide range of ionization.

We included both dust-free and dusty simulations. Our dust-free models are more successful at reproducing the observations, arguing that star forming galaxies on average have lower dust to gas ratios than the Orion Nebula composition that we used in our dusty models. Our main conclusions in this paper are unaffected by whether or not dust is included because the primary diagnostic ratios are relatively insensitive to the effects of dust.

Our models show that the one-dimensional sequence from high to low ionization objects picked out by MFICA, the "SF locus," represents a physically meaningful sequence best interpreted by a changing relative concentration towards higher values in the distribution of the ionizing flux $\phi_H$ incident on the individual gas clouds scattered throughout a complicated SF galaxy. The sense of this relationship is that high ionization galaxies will have distributions of $\phi_H$ weighted towards greater values of ionizing flux incident on the optimally emitting clouds than their lower ionization counterparts. A similar relationship was previously postulated by M01. Future work could confirm this relationship between ionization level and distribution of $\phi_H$ by investigating lower redshift SDSS samples, and by determining a suitable set of spatially resolved narrow-line-emitting starburst galaxies.

The fits of our LOC models to the observed spectra are somewhat improved by also varying the metallicity, but metallicity changes alone do not provide nearly the full range in ionization level that is observed. A shift from dust-free models for the highly ionized galaxies to dusty models for the low-ionization galaxies also improves the fits. However, in either of these cases it is still systematic changes in the distribution of the ionizing flux striking the gas that provides most of the variation in the observed spectra along the SF sequence.

This work shows that star forming galaxies can be modeled using the same methodology as AGN with locally optimally emitting clouds. The simplicity of this method for understanding galaxies over the entire span of the BPT diagram, excluding LINERs, is an important step. Compared to earlier models of SF galaxies such as those by Kewley et al. (2001) and L10, we have replaced the use of a single ionization parameter and constant gas density (implying a constant ionizing flux) to describe all gas clouds in a particular galaxy with a more realistic description in which gas clouds scattered throughout a galaxy see a wide range in ionizing flux and have a wide range of gas densities. As such, this represents substantial progress in physically interpreting emission line galaxies as a whole.

The greatest strength of this work comes from presenting an alternative interpretation to the "abundance sequence" originally developed by Kewley et al. (2001) and later expanded upon by L10. Here, we have shown that differences in metallicity between low- and high-ionization starburst galaxies might not be the primary driver for the systematic variation in their emission line properties. Instead, the trends among such galaxies could largely be caused by variations in the distribution of the ionizing flux $\phi_H$ incident on the line-emitting gas clouds, with the higher-ionization galaxies having a distribution in $\phi_H$ more heavily weighted towards larger $\phi_H$ values. A possible underlying cause might be age effects as the individual H II regions expand. This new interpretation potentially affects results based on the idea that metallicity and ionization are correlated over the entire SF wing of the BPT diagram, therefore requiring significant reevaluation about the underlying nature of SF galaxies.

**ACKNOWLEDGEMENTS**

CTR wishes to acknowledge the support of the Extreme Science and Engineering Discovery Environment (XSEDE), which is supported by National Science Foundation grant number OCI-1053575, and Elon University's FR&D Summer Research Fellowship. The Michigan State University High Performance Computing Center (HPCC) and the Institute for Cyber Enabled Research (ICER) also supported this work. CTR and JAB acknowledge NSF support for this work under grant AST-1006593. CTR and HM wish to acknowledge the support of Elon University's Lumen Prize, Summer Undergraduate Research Experience (SURE), and Honors Program. JTA acknowledges the award of a SIEF John Stocker Fellowship. PCH acknowledges the support of the UK Science and Technology Research Council (STFC). GJF acknowledges support by NSF (1108928, 1109061, and 1412155), NASA (10-ATP10-0053, 10-ADAP10-0073, NNX12AH73G, and ATP13-0153), and STScI (HST-AR- 13245, GO-12560, HST-GO-12309, GO-13310.002-A, and HST-AR-13914). AC acknowledges Elon University for his sabbatical leave.